\documentclass[11pt]{article}

\usepackage{footnote}

\usepackage[utf8]{inputenc}
\usepackage{amsfonts,url,epsfig}
\usepackage{geometry} 
\usepackage{sectsty} 
\usepackage[normalem]{ulem}

\usepackage{nohyperref}
\usepackage{todonotes}
\usepackage[super,sort&compress]{natbib}
\setcitestyle{comma,numbers,super,open={},close={}} 
\makeatletter
\renewcommand\@biblabel[1]{#1.}
\makeatother

\sectionfont{\sffamily\bfseries\upshape\large}
\subsectionfont{\sffamily\bfseries\upshape\normalsize} 
\subsubsectionfont{\sffamily\mdseries\upshape\normalsize}
\makeatletter
\renewcommand\@seccntformat[1]{\csname the#1\endcsname.\quad}
\def\@maketitle{%
  \begin{center}%
  \let \footnote \thanks
    {\large \@title \par}%
    {\normalsize
      \begin{tabular}[t]{c}%
        \@author
      \end{tabular}\par}%
    {\small \@date}%
  \end{center}%
}
\makeatother

\title{\bf Slamming the sham:\\A Bayesian model for adaptive adjustment with
noisy control data\footnote{We thank the U.S. Office of Naval Research, Institute for Education Sciences, Sloan Foundation, and 
NWO Veni grant number VI.Veni.202.124 for
partial support of this work.  Data and code  are at
\href{https://github.com/VMatthijs/Slamming-the-sham}
{https://github.com/VMatthijs/Slamming-the-sham} }\vspace{.1in}}

\author{Andrew Gelman\footnote{Department of Statistics and Department of
Political Science, Columbia University, New York.} \ and Matthijs
Vákár\footnote{Department of Information and Computing Sciences, Utrecht University, Utrecht.} 
\vspace{.1in}}

\date{26 Feb 2021\vspace{-.2in}}

\begin{document}
\maketitle

\begin{abstract}
  It is not always clear how to adjust for control data in causal inference,
  balancing the goals of reducing bias and variance.  We show how, in a setting with repeated
  experiments, Bayesian hierarchical modeling yields an adaptive procedure that
  uses the data to determine how much adjustment to perform.
  The result is a novel analysis with increased statistical efficiency compared to the 
  default analysis based on difference estimates.
  We demonstrate
  this procedure on two real examples,
  as well as on a series of simulated datasets.
  We show that the increased efficiency can have real-world consequences in terms of the conclusions
  that can be drawn from the experiments.
  We also discuss the relevance
  of this work to causal inference and statistical design and analysis more
  generally.
\end{abstract}

\section{Introduction}
Consider the following problem.  Experiments
$j=1,\dots,J$ are performed, and each is paired with a sham experiment with a
null treatment.  Label the estimated treatment effects for each experiment $j$
as $y_{j1}$ for the active data and $y_{j0}$ for the sham data.  It is standard
practice to estimate the treatment effect in experiment $j$ as $y_{j1} -
y_{j0}$.  But this bias adjustment can add noise.  In many cases, it is not a priori obvious
whether the sham experiments can be safely discarded or not.
The same problem arises in observational studies in economics with the difference-in-differences estimate (see, e.g., Ashenfelter, Zimmerman, and Levine\cite{AshenfelterZimmermanLevine2003}), where, again, subtracting the baseline difference can reduce bias but at the cost of increasing variance.

\textit{How can we decide whether to adjust for the sham data and how best to do so, if we do adjust?}  We propose a hierarchical Bayesian approach which is broadly consistent with modern ideas of regularization in causal inference for varying treatment effects (e.g., Hill\cite{Hill2011} and Wager and Athey\cite{WagerAthey2018}).  We move beyond standard Bayesian meta-analysis (e.g., Smith, Spiegelhalter, and Thomas\cite{SmithSpiegelhalterThomas1995} and Higgins and Whitehead\cite{HigginsWhitehead1996}) by partially pooling biases as well as the treatment effects, thus allowing an adjustment that adapts to observed variation in the sham data.  Our approach uses Bayesian multilevel modeling and is most generally effective when the number of experiments, $J$, is large.  When $J$ is small, strong prior information is required to most efficiently use the sham information; when $J$ is large, the relevant hyperparameters can be estimated from the data.

The core contributions of this paper are:  (1) a focus on a problem that arises in many areas of science when analyzing repeated controlled experiments, and where an existing default method can yield demonstrably poor performance, (2) a solution, along with code to implement it, (3) a method for interpreting the resulting estimate as an approximate partial adjustment, and (4) a set of simulation-based evaluations of the method that are directly relevant to the ways in which these studies are reported.  

We demonstrate the need for a solution to the sham-adjustment problem, and our recommended method, in the context of two real examples. First, we consider a series of laboratory experiments on the effects of electromagnetic fields on calcium flow in the brain. The results of these experiments were influential in a public health debate regarding cancer clusters that had been found near electric power lines.  For the experimental results under study, it possible to greatly improve the published analysis by modeling the bias rather than simply subtracting the sham estimate, and the new analysis alters the scientific conclusions. As a second case study, we consider a recent, highly cited meta-analysis on repetitive transcranial magnetic stimulation as a treatment for depression by  Berlim et al.\cite{Berlim} \ Our method addresses a problem that is widely found across science, however, whenever repeated controlled experiments are analyzed, and we could have equally have considered the studies of Fuchikami et al.\cite{Fuchikami}, Kádár et al.\cite{KadarELimCarrerasGenisDTemelHuguet2011}, or Le Quément et al.\cite{Quement}, to name a few.

\section{The problem and proposed solution}

\subsection{The model and two estimates}

Suppose that, for each of $j=1,\dots,J$, two experiments have been conducted, yielding estimate $y_{j1}$  and standard error $s_{j1}$ from the active-treatment experiment and $y_{j0}$ and $s_{j0}$ from the sham-treatment experiment.  We consider the following model that is intended to capture the experiment and estimation process for each pair of experiments $j$, where we assume statistical independence between all pairs of experiments:
\begin{eqnarray}
\nonumber  && y_{j1} \sim \mbox{normal}(\theta_j + b_j, s_{j1})\\
 \label{2equations}  && y_{j0} \sim \mbox{normal}( b_j, s_{j0}).
\end{eqnarray}
Here, $\theta_j$ is the treatment effect of interest and $b_j$ is an experimental bias
shared by the real and sham treatments.  In modeling the bias in this way we are following the general approach of Greenland\cite{Greenland2005}.  For simplicity of presentation and for application to the meta-analysis problem, we shall assume that the estimates and standard errors are given, and that the sample size in each experiment is large enough that it is reasonable to approximate the information from the data in the form of normal likelihoods with known variances.  It would not materially affect the methods or conclusions of this paper if we were to go to the raw data (where available) or to replace the normal with $t$ likelihoods corresponding to the degrees of freedom of the data in each experiment.

We start by considering two estimates of the treatment effect $\theta_j$:  the exposed-only estimate, $y_{j1}$, and the difference estimate, $y_{j1} - y_{j0}$.

Under model (\ref{2equations}), the difference estimate is unbiased---indeed, it is the only unbiased estimate of $\theta_j$.  However, performing this subtraction adds noise, doubling the variance if the standard errors of the active treatment data and sham are the same.  If the bias $b_j$ in the experiments were zero, the exposed-only estimate would clearly be the better choice.  More generally, depending on the size of the bias, $b_j$, it could be more effective to partially adjust for the sham rather than to fully subtract $y_{j0}$.

At this point, a scientist might feel that the safe choice would be to use the difference estimate, paying the price of a higher mean squared error, as it could seem risky to accept bias.  Researchers are often trained to think of bias as the primary concern, with the minimum-variance unbiased estimator being optimal\cite{LehmannScheffe1950}.  We suspect that such an attitude is not as prevalent as in the past, now that we are used to regularization in methods ranging from lasso to deep learning to multilevel regression, but it remains a starting point in many analyses.

In the present paper we shall consider the exposed-only and difference estimates as two extreme cases of a Bayesian procedure that performs meta-analysis on the treatment effects and biases.  We first present the Bayesian model, then demonstrate its merits on the applied example that motivated this research as well as on a more recent example that illustrates different aspects of the model, and then present methods for understanding and evaluating the inferences.

\subsection{Multilevel model and Bayesian analysis}\label{bayes}

We have set up model (\ref{2equations}) in a way to reflect the scientific
choices indicated in design and data collection. 
The next step is the model for the treatment effects and the biases.  This is
the multilevel part of the model, and by default we will use normal
distributions (again, assuming independence across the different values of $j$):
\begin{eqnarray}
  \nonumber  b_j &\sim&\mbox{normal} (\mu^{b}, \sigma^{b} )\\
    \label{meas}    \theta_j &\sim&\mbox{normal} (\mu^{\theta}, \sigma^{\theta} ).
\end{eqnarray}
We briefly go through the hyperparameters of this model:
\begin{itemize}
  \item $\mu^{\theta}, \sigma^{\theta}$ are the mean and standard deviation of
        the true effects.  $\mu^{\theta}$ and
        $\sigma^{\theta}$ determine the partial pooling in the estimates of the
        individual $\theta_j$'s.
  \item $\mu^b$ is the average experimental bias and will equal zero if the sham
        treatments have no effect.
  \item $\sigma^b$ is the variation in the biases across experiments and, again,
        will equal zero if the sham treatments have no effect.
\end{itemize}
We need to include an average sham effect and variation in the sham effects in
the model to allow for the possibility of bias.  This is a matter of respecting
the experimental design:  the sham treatments were included in the study for a
reason.

We can fit the model using Bayesian inference with default uniform priors on the
hyperparameters $\mu^{\theta}, \sigma^{\theta},\mu^b,\sigma^b$, with the
understanding that informative priors could be used in problems where such prior
information is readily available. We choose a Bayesian approach (rather than
using marginal maximum likelihood to obtain a point estimate of the
hyperparameters) because it accounts for the uncertainty in the hyperparameters,
and also for computational convenience---we can fit our model directly in Stan\cite{carpenter2017stan}, and it is easy to extend the Stan model to include departures from normality, linearity, and exchangeability as desired.

Model (\ref{meas}) represents a default, or starting point.  In real-world meta-analyses there can be additional prior information, and the $J$ studies will differ in various known ways.  Suppose we have a predictor $x_j$ assigned or observed for each study, $j$.  Then it will make sense to allow the expected treatment effect to vary by $x$, thus replacing the exchangeable model for $\theta$ in (\ref{meas}) by something like,
\begin{equation}
  \label{meas2}  \theta_j = g(x_j),
\end{equation}
where $g$ is a stochastic function whose distribution will itself depend on hyperparameters, for example a linear regression with errors, $g(x_j) \sim \mbox{normal}(a + bx_j, \sigma^{\theta})$, or a Gaussian process that penalizes discrepancies between $g(x_j)$ and $g(x_k)$ for nearby pairs $(x_j,x_k)$. The choice of model for $g$ will depend on the particular applied problem.

It would also be possible to add structure to the model for the biases $b_j$. For example, a correlation between $b_j$ and $\theta_j$ would allow biases to be larger under conditions of larger treatment effects, which could make sense in some contexts.

\subsection{Frequency evaluation}\label{freq}
In applied statistics it is not enough to come up with a good estimate; it is also necessary to understand it and compare to previously existing approches.  To this end, we compare the estimated treatment effects under the hierarchical model to the exposed-only estimates, $y_{j1}$, and the difference estimates, $y_{j1} - y_{j0}$.  We conduct this evaluation using a simulation study, as we demonstrate in Section \ref{evaluation} for our motivating example.  The simulation study is conducted to allow a range of values for the crucial parameter $\sigma^b$ which governs the value of the information from the sham experiments.

For each of the three estimates, we then compute the following four summaries:  (i) the proportion of the $J$ estimates that are statistically
significant (that is, where the estimate $\pm 1.96$ standard error
or Bayesian 95\% posterior interval excludes zero), (ii) the type S error rate
(the proportion of statistically significant estimates that are the
wrong sign), (iii) the mean squared error of the $J$ estimates compared the true
values $\theta_j$ (which by the design of the simulation are known to us), and
(iv) the correlation between the ranks of the $J$ estimates and the ranks of the
true $\theta_j$'s.

We choose these summaries because they represent four different practical goals
of this sort of study:  (i) identification of experiments where the treatment
effect is statistically distinguishable from zero, (ii) validity of
these claims of confidence, (iii) accurate estimation of treatment effects, and
(iv) ranking of which results are strongest and most worthy of further study.  It is important in any frequency evaluation to consider statistical properties that are relevant to the task at hand, and we argue for the relevance of these measures in the context of our applied example.

\subsection{Relevance and novelty of this procedure}

This model can apply to a large set of problems of repeated controlled experiments, such as arise in biology, medicine, policy analysis, and other fields where a treatment effect is conjectured to vary in some unknown way as a function of input conditions, so that the point of the study is not merely to estimate an average treatment effect but also to estimate the individual $\theta_j$'s.  In Section \ref{chicks}, we consider an example from biology in which the goal was to estimate the dependence of $\theta$ on $x$; in Section \ref{berlim} we consider a medical example where the distribution of the $\theta_j$'s was of interest.

The hierarchical model and Bayesian computation used in this paper are now familiar statistical tools.  What is new here is, first, their application to a causal inference setting where it is often standard practice to simply subtract sham estimates (sometimes called a difference-in-difference procedure) rather than to jointly model active and sham data; second, the frequency evaluation demonstrating the superiority of the modeling approach under a wide range of conditions; and, third, the expression of the Bayesian estimate as an approximate fractional adjustment for the sham, which links these results to existing practice.

\section{Applied example 1:  Magnetic fields and calcium efflux}\label{chicks}

\subsection{Background}
The 1980s saw a concern regarding health effects of low-frequency
magnetic fields, as a result of some findings in epidemiology that children
living near electric power lines had elevated risks of leukemia, and this caught the interest of the news media\cite{Brodeur1989a,Brodeur1989b,Brodeur2000}. One posited
mechanism for a carcinogenic effect here was that magnetic
fields interfered with cell structure, and this general model was studied in a
series of experiments conducted at the U.S. Environmental Protection Agency,
measuring the effects on calcium efflux in chick brains. The studies were
carefully conducted with an eye toward theory, measurement, and statistical
design\cite{Blackman2015}.  Each chick brain was divided in two, with one half of
the brain randomly assigned to the treatment of exposure to an alternating current magnetic field
at a specified frequency and the other brain half given the control of no
exposure to the field.  Between 28 and 36 chicks were employed in each
experiment, and 38 experiments were performed, representing magnetic field
frequencies ranging from 1 to 510 Hz; see Blackman et al.\cite{BlackmanBenaneElliottHousePollock1988}

\begin{table}
  \begin{small}
    \centerline{
      \begin{tabular}{c cc cc}
                       & \multicolumn{2}{c}{Sham treatment} &
        \multicolumn{2}{c}{Real exposure} \\
        Frequency (Hz) & $n$                               & Estimate $y_{j0}$ (s.e. $s_{j0}$)
        & $n$ & Estimate $y_{j1}$ (s.e. $s_{j1}$) \\ \hline
        \ 1            & 32                                & $-0.005$ (0.041)
        & 32  & 0.036  (0.041)  \\
        15             & 32                                & $\ \ 0.013$
        (0.042)                & 36  & 0.173 (0.034)   \\
        30             & 32                                & $\ \ 0.033$ (0.032)
        & 32  & 0.107  (0.035)  \\
        45             & 32                                & $ -0.010$  (0.032)
        & 32  & 0.181  (0.052) \\
        \dots & \dots & \dots & \dots & \dots
      \end{tabular}
    }
  \end{small}
  \caption{\em A portion of the data summaries from the chick brains experiment
    reported by Blackman et al.\cite{BlackmanBenaneElliottHousePollock1988}  Data continue at 15 Hz intervals all
    the way up through 510 Hz.  As can be seen from the above numbers, the sham
    estimates are statistically indistinguishable from zero, whereas the effects
    are clearly positive for many of the real experiments.}
  \label{rawdata}
\end{table}

As a check against systematic bias, each experiment was repeated under ``sham''
conditions, with the same setup but with the magnetic field turned off.  Each
sham and real experiment was then analyzed to produce an estimated relative
effect, along with a standard error. The experimental design also included
clustering, but we do not further consider that here. Unfortunately the authors
refused to share their data when requested, and so in our
analysis we are restricted to the published data summaries, which are the
estimates and standard errors for each sham and real experiment. A subset of
these data summaries are displayed for clarity in Table \ref{rawdata}.

In the published analysis, the effect of magnetic fields at each frequency was
estimated by subtracting the estimates from the real and sham exposures, adding
the variances as is appropriate for independent experiments.  The estimate for
each experiment $j$ is then $y_{j1} - y_{j0}$, with standard error
$(\sigma^2_{j1} + \sigma^2_{j0})^{1/2}$.

Is it appropriate to subtract the sham estimate?  An alternative would be to
simply use the estimate from the real exposure, $y_{j1}$ with its standard
error, $\sigma_{j1}$, which discards the sham data entirely and has the benefit
of having approximately half the variance of the differenced estimator.

The difference, $y_{j1} - y_{j0}$, would typically be considered a safe and
conservative estimate as it corrects for any biases shared by the two
experiments, and it indeed was used in
the published paper and not questioned in that literature.  However, as we shall
see in our discussion of the inferences and conclusions drawn from these data,
reliance on the noisy differenced estimator may well incur real scientific
costs.

\subsection{Originally published analysis}\label{orig}

\begin{figure}[h]
  \centerline{
    \includegraphics[width=.5\textwidth]{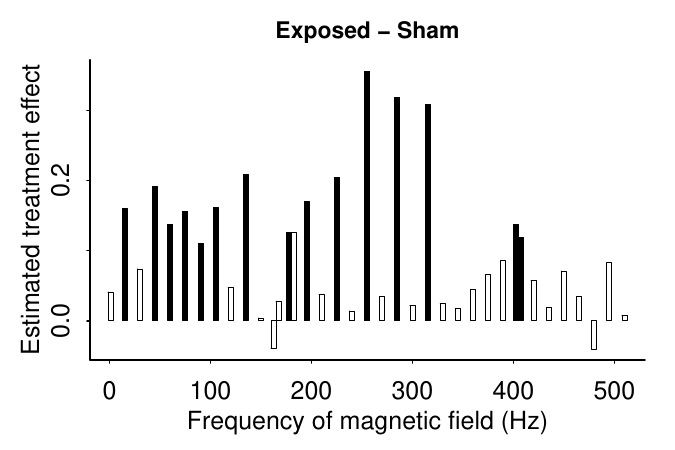}
    \includegraphics[width=.5\textwidth]{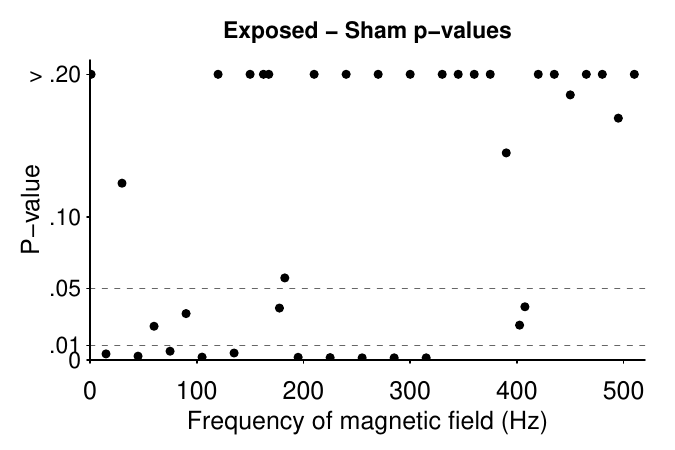}}
  \vspace{.1in}
  \centerline{
    \includegraphics[width=.5\textwidth]{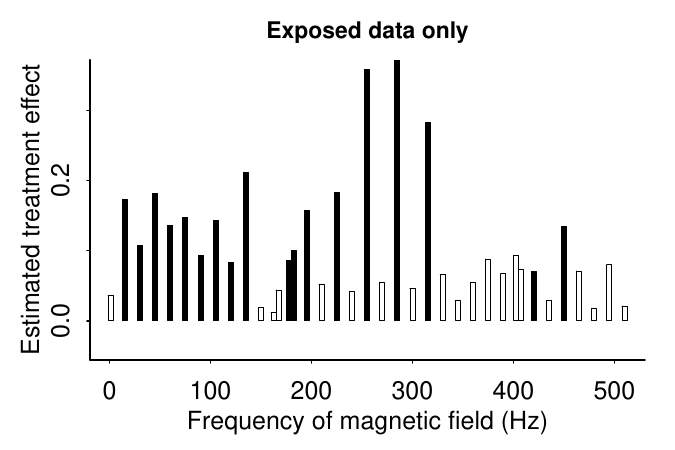}
    \includegraphics[width=.5\textwidth]{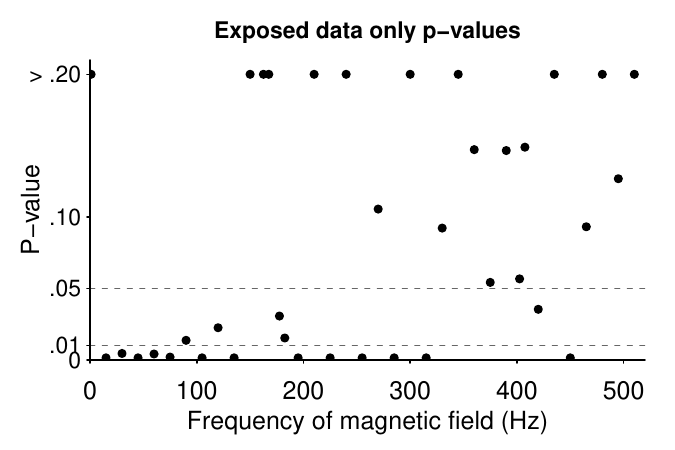}}
  \vspace{-.1in}
  \caption{\em {\em Top row:}  Redrawn versions of the graphs of  Blackman et
    al.\cite{BlackmanBenaneElliottHousePollock1988}, summarizing the chick brains data by categorizing estimates at
    different frequencies based on their statistical significance:  (a)
    Estimates $y_{j1} - y_{j0}$ plotted vs.\ frequencies $x_j$.  For three of
    the frequencies (165, 180, and 405 Hz) the experiment was performed twice,
    and in these cases we have jittered the two experimental results so they
    both appear on the graph.  Each bar is shaded if the experimental result is
    statistically significant at the 5\% level based on the appropriate $t$
    distribution.  (b) Results of each experiment displayed as a
    $p$-value.\newline
    {\em Bottom row:}  Corresponding plots using only the exposed data,
    $y_{j1}$.  The patterns are similar but with enough differences to change
    some of the reported results.}
  \label{redrawn}
\end{figure}

 Blackman et al.\cite{BlackmanBenaneElliottHousePollock1988} presented the differenced estimates and categorized them
based on levels of statistical significance relative to the hypothesis of zero
effects. The top row of Figure \ref{redrawn}a shows redrawn versions of the
graphs in that paper.  The top-left graph displays point estimates, shading
those that are statistically significant.  The top-right graph shows $p$-values
of the hypothesis of zero effect at each frequency.\footnote{Our Figure
\ref{redrawn}b is slightly different from Figure 2 of  Blackman et al.\cite{BlackmanBenaneElliottHousePollock1988}
for reasons that are not clear to us, as our displayed $p$-values are consistent
with those in Table 1 of Blackman et al.\cite{BlackmanBenaneElliottHousePollock1988}, but in any case the differences are
minor and do not affect the arguments of this paper. It was not possible to obtain the raw data from these experiments} The authors divide these
into three categories:  those with $p$-values less than 0.01, those with
$p$-values between 0.01 and 0.05, and the rest.

This division based on statistical significance was a mistake, and it is a
common mistake in applied statistics; see Gelman and Stern\cite{GelmanStern2006}.  Seemingly
major differences in $p$-values are not necessarily statistically significant or
even close to significant.  For example, $p$-values of 0.20 and 0.01 correspond
to $z$-scores of 1.28 and 2.33, respectively (using the normal distribution here
for simplicity).  So, even though $p=0.20$ seems like no evidence at all, while
$p=0.01$ appears to be a very strong result, their difference is a mere 1.05
standard errors, which can easily occur by chance.

The use of a $p$-value-based decision rule had consequences.  In the paper under
discussion,  Blackman et al.\cite{BlackmanBenaneElliottHousePollock1988} used the summary shown in the top-right
graph of Figure \ref{redrawn} to draw the following conclusions: ``those data
with $P$-values less than 0.01, which extend from 15 to 315 Hz, could form one
set composed of two groups of 30 Hz \dots the response at 60, 90 and 180 Hz, the
first odd multiple of 60 Hz, with an elevated but not statistically reliable
response at 30 Hz, may be part of a second set \dots the response at 405 Hz may
represent still another set \dots.''  To their credit, the authors emphasized
that these are ``only hypothetical constructs,'' but these noisy results formed
the empirical conclusions of the paper and they motivated in the published paper a further three-page
speculation about physical models.

The specific claims from these and similar experiments also influenced perceptions of researchers and the public regarding underlying mechanisms.  For example, in an article published in a respected national magazine,  Brodeur\cite{Brodeur1989b} reported,  ``Blackman was trying to figure out why fields with frequencies of fifteen, forty-five, seventy-five, and a hundred and five hertz should have such a strong effect on calcium-ion outflow from chick-brain tissue, while fields of thirty, sixty, and ninety hertz produced only a weak effect.''

\subsection{Exploration of the sham data}

\begin{figure}
  \centerline{
    \includegraphics[width=.5\textwidth]{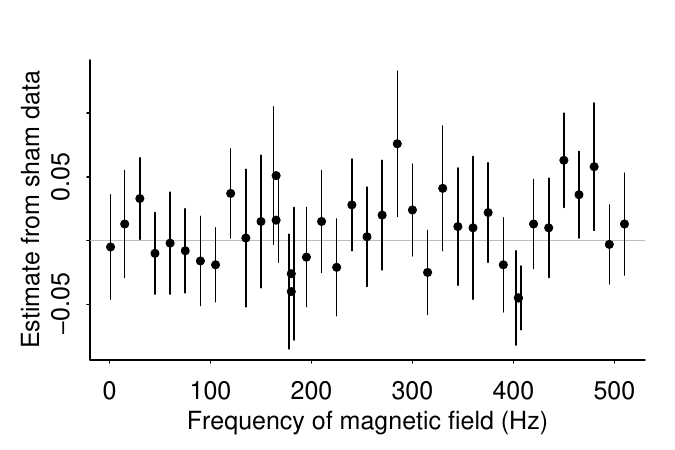}
    \includegraphics[width=.5\textwidth]{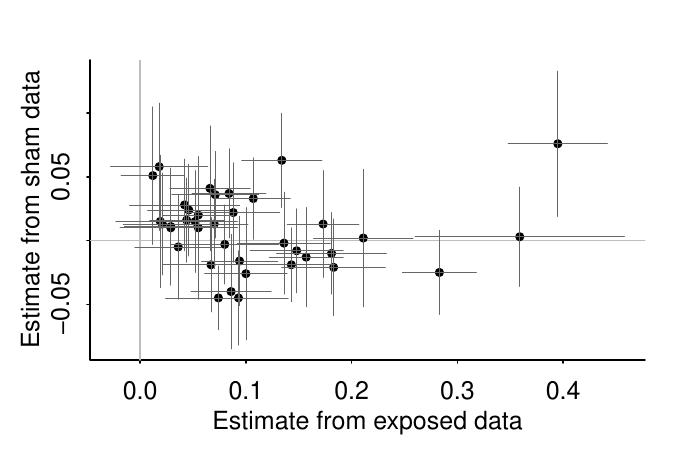}
  }
  \vspace{-.1in}
  \caption{\em (a) Estimates $\pm$ standard errors of the effect of the sham
    treatment as a function of frequency of the (turned off) electromagnetic
    field; (b) Sham vs.\ exposed estimates.  Unsurprisingly, given the careful
    design of the experiment, there is no evidence that the sham effects are
    anything other than zero.}
  \label{sham}
\end{figure}

Let us set aside concerns about summarizing experimental results by
discretizing $p$-values, an approach that has been increasingly contested in
recent years\cite{WassersteinLazar2016}, and instead focus on the question of
what should be done with the data from the sham experiments in the chick study.

A glance at Table \ref{rawdata} suggests that nothing much seems to be going on
in the sham data, which is confirmed by examination of the entire dataset:  the
estimates fluctuate around the zero, with the amount of variation consistent
with the reported standard errors; see Figure \ref{sham}a. This impression can
be confirmed with a simple $\chi^2$ test:
$\sum_{j=1}^{38}(y_{j0}/\sigma_{j0})^2= 21.3$, which is quite a bit {\em less}
than would be expected under the $\chi^2_{38}$ distribution.  This suggests
there may be an problem with the standard errors, as they seem to be too
conservative---perhaps there was an error in their computation, as the data were
collected using a clustered design and perhaps this was not correctly handled in
the standard error calculations---but, in any case, there is no evidence for any
variation in the effects of the sham treatment. Furthermore, the mean of the 38
sham estimates is 0.01, which is both substantively and statistically
insignificantly different from the null, so the data do not contradict the model
of no sham effect.  This should be no surprise---given that the magnetic fields were turned off in the sham condition, we would not expect a null treatment to have
any effect, and the sham experiments represent an abundance of caution more than
anything else.

In our remaining treatment of these data we shall take the sham estimates and
standard errors as reported; arguably, though, it would make sense to scale all
the standard errors down by a factor of $\sqrt{21.3/38}$ as an approximation to
the adjustment that would be required, under the assumption that some mistake
was made in their calculation.  Scaling these standard errors down would not
affect our main conclusions; indeed it would just make our advocacy of an
alternative analysis even stronger by increasing the precision of our
inferences.

To continue with our main thread, in Figure \ref{sham}b we look for patterns in
the sham data another way, by plotting the sham estimate $y_{j0}$ vs.\ the
exposed estimate $y_{j1}$ for each frequency $j$. We see no pattern, which again
is consistent with the sham estimates being pure noise.

\subsection{Analysis not adjusting for the sham data}

If the sham estimates are indeed nothing but noise, then it makes sense not to
include them in the estimated treatment effects.  The resulting unadjusted
analysis is simple:  just report $y_{j1}$ with standard error $\sigma_{j1}$ at
each frequency $j$. We could almost describe this as ``analysis ignoring the
sham data'' but that would not quite be correct.  We did not ignore the sham
data:  we only decided to exclude the sham data from our inferences after first
analyzing the sham results and finding no evidence distinguishing them from pure
noise.

The bottom row of Figure \ref{redrawn} shows the results.  We use the same sorts
of displays as used in the earlier published paper, not because we think it
appropriate to summarize a set of experiments using statistical significance but
because we wish to demonstrate the potential practical gains that could come
from switching to the undifferenced estimates, even without considering
alternative inferential summaries.

For this example, it is clear from a modern perspective that the
estimates $y_{j1}-y_{j0}$ are inferior to the simple $y_{j1}$.  The sham
experiments may well have been an important part of the design of the study, as
they rule out a potential threat to validity in the causal inferences, but given
what the data look like, it is not necessary to include their data in the final
estimates.

The challenges we address in this paper are, first, to come to this conclusion
in a more systematic way; second, to situate this in a general framework that
can apply to other designs; third, to come up with a compromise solution for
settings where the sham data are noisy but contain some information; and,
fourth, to be able to report such a compromise estimate in a reasonable way.

\subsection{Scientific consequences of the choice of analysis}

We now go through the original conclusions drawn from the chick study and see
how they could have differed, had they been based on the bottom row of Figure
\ref{redrawn} rather than the more noisy, statistically inefficient summaries
shown in the top row of that figure.

Perhaps most importantly, the overall impression of the data would have changed.
 Blackman et al.\cite{BlackmanBenaneElliottHousePollock1988} started off by declaring: ``These results demonstrate
that certain frequencies are effective ($P<.05$) in causing enhance calcium-ion
efflux while others are not.''  And, indeed, the upper-left plot of Figure
\ref{redrawn} shows a mix of positive and negative results, and most are not
statistically significant.  In contrast, in the lower-left plot all the point
estimates are positive, making it clear that the results are consistent with a
general pattern of positive effects with uncertainty at individual frequencies.

Removing the sham correction affects more detailed conclusions as well.  Blackman
et al.\cite{BlackmanBenaneElliottHousePollock1988} pulled out patterns from the top-right graph Figure \ref{redrawn}
that do not appear when this same $p$-value classification is used in the 
bottom-right graph.  They labeled one set of responses as occurring at five
frequencies at the low end---15, 45, 75, 105, and 135 Hz---but in the new graph
the frequencies of 30 and 60 Hz also fall in this $p<0.01$ category, destroying
the alternating pattern of positive and null results.  Relatedly, Blackman et al.\cite{BlackmanBenaneElliottHousePollock1988} placed 60,
90, and 180 Hz in together in a set of intermediate $p$-values---but in the
cleaner summary, this category contains 120 Hz rather than 60 Hz, obviating a
discussion later in the paper of how ``the data at 180 Hz could be the
fundamental of a nonlinear mechanism \dots leading to subharmonic frequencies
that manifest at 90 and 60 Hz.''

The article also included speculation about what was going on at 405 Hz, which in
the original analysis was the only frequency at the high end with a statistically
significant effect; see the top-left graph of Figure \ref{redrawn}.  The
revised, bottom-left, plot tells a completely different story:  the estimate at
405 Hz is no longer statistically significant, but those at 420 and 450 Hz are.
An entirely new set of theories would be needed to explain this pattern.

We are not saying that it was a bad idea for the authors of the original paper
to engage in data-based scientific speculation.  Rather, our point is that the
statistically inefficient decision to adjust for the sham data is not merely of
theoretical interest; it has real effects on the empirical conclusions from this
study and also on the scientific explanations proposed for further study.  The
analysis subtracting the sham estimates may have seemed at the time like a safe
choice, but in this example it simply added noise.

For an example of the practical impact of not fully modeling variation, consider this quote from  Blakeslee\cite{Blakeslee1991}:  ``This requirement for exact field geometries may help explain the `Cheshire cat phenomenon' in bioelectromagnetic experiments, Dr.\ Blackman said. Researchers have long been vexed by a now-you-see-it, now-you-don't problem as many experiments were not reproducible from one laboratory to the next, he said.''  It seems that active research effort was devoted to studying experimental differences which well may have been explainable by noise.  We consider this not a criticism of these particular researchers so much as a general concern with the routine use of simple statistical analyses (in this case, subtraction of the sham estimate) which lead to unnecessarily variable conclusions.

\subsection{Reanalysis using the multilevel model}\label{posterior.fit}

\begin{table}
  \begin{small}
    \centerline{
      \begin{tabular}{ccc}
        Parameter          & Estimate (s.e.) & 95\% interval        \\\hline
        $\mu^{\theta}$     & 0.097 (0.015)   & $\ \ [0.069, 0.126]$ \\
        $\sigma^{\theta} $ & 0.069 (0.014)   & $\ \ [0.044, 0.099]$ \\
        $\mu^{b}$          & 0.004 (0.006)   & $[-0.008, 0.017]$    \\
        $\sigma^b$         & 0.008 (0.006)   & $\ \ [0.000, 0.021]$ \\
      \end{tabular}
    }
  \end{small}
  \caption{\em Posterior means, standard deviations, and 95\% intervals  for the
    hyperparameters in the hierarchical model fit to the chick
    data.}\label{inferences}
\end{table}

We now fit the multilevel model (\ref{meas}) to the Blackman et al.\ data; inferences for the
hyperparameters appear in Table \ref{inferences}.  The estimates of
$\mu^{\theta}$ and  $\sigma^{\theta}$ imply a distribution of treatment effects
with a clearly positive mean, along with substantial variation, implying
different effects at different frequencies.  But there is no evidence for any
sham effects:  both $\mu^b$ and $\sigma^b$ are estimated to be essentially
zero---even at the highest end of the uncertainty interval, a value of 0.02
would be a tiny amount of bias compared to treatment effects that are three to
six times higher. The lack of evidence for any sham effects is no surprise given
the preliminary analysis shown in Figure \ref{sham}.  Again, the point of our
hierarchical model in this example is not to discover the evident lack of
noticeable sham effects but rather to be part of a general approach to this sort
of problem.

\begin{figure}
  \centerline{
    \includegraphics[width=.5\textwidth]{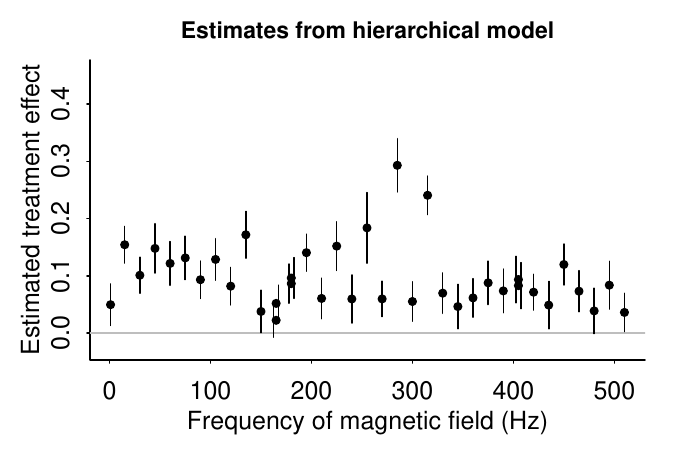}
    \includegraphics[width=.5\textwidth]{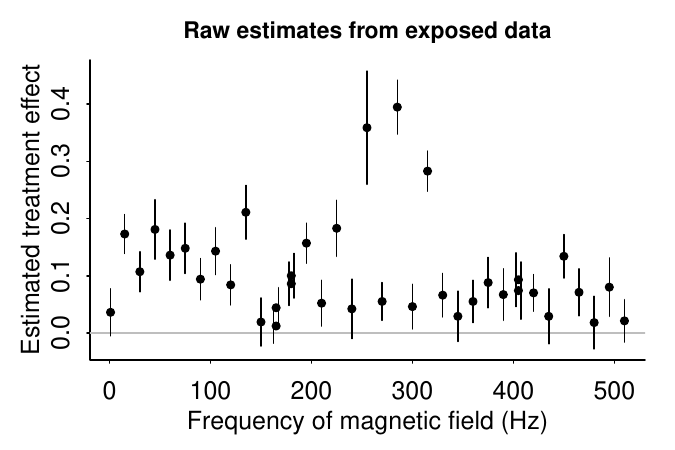}}
  \vspace{-.1in}
  \caption{\em  (a) Posterior mean $\pm$ standard deviation of each treatment
    effect $\theta_j$ from the hierarchical model fit to the chick data.  The
    fitted model estimated the sham effects to be essentially zero (see
    discussion of Table \ref{inferences}), and so these estimated treatment
    effects come pretty much from the exposed data alone. For three of the
    frequencies (165, 180, and 405 Hz) the experiment was performed twice, and
    in these cases we have jittered the two experimental results so they both
    appear on the graph.  (b) For comparison, the raw estimates $y_{j1}$ from
    the exposed data $\pm$ standard error.  The Bayesian hierarchical estimates on the left
    plot are partially pooled toward a common mean.}
  \label{hierarchical}
\end{figure}

Figure \ref{hierarchical}a shows the posterior mean and standard deviation of
the treatment effect $\theta_j$ for each experiment $j$.  For comparison, we
display in Figure \ref{hierarchical}b the raw estimates $y_{j1}\pm\sigma_{j1}$
from the exposed data.  The estimates from the hierarchical model have been
partially pooled toward the common mean but otherwise show a pattern similar to
that of the raw data, with the largest change being the raw estimate at 255 Hz
that had a very large standard error (the long error bar in Figure
\ref{hierarchical}b) and was thus pulled closer to the center of the
distribution.

Again, we are not surprised that our Bayesian inferences are
qualitatively similar to the raw estimates from the exposed data.  Recall that
this whole example came up because the standard recommendation to subtract the
sham data yielded unnecessarily noisy estimates.  We consider it a success that
hierarchical modeling gives us a general approach that also arrives at a reasonable
conclusion in this particular case.

In the data at hand we see no clear patterns or correlations that would warrant a more structured model for the treatment effects or the biases; however, in Appendix \ref{alt-mod} we consider some alternative models of the form (\ref{meas2}), as a robustness check and also to demonstrate how such models could be fit using Stan.

\subsection{How common is the situation of zero sham effects?}
In this example, the sham effects were estimated to be essentially zero.
A reader might wonder how rare this situation is and whether our example generalizes.
We argue that this situation is, in fact, quite common.

In the chick brains example, effect sizes are frequently quantified as a ratio between 
a pre- and post-treatment measurement, both for the sham and real treatments. 
In case of a sham treatment without effect, this ratio tends to be 1.
Either subtracting 1 from both treatment and sham effects or taking their logarithm
leaves us with a measure of sham effects that is expected to be 0 in case of a truly 
irrelevant sham treatment.  More generally, controls are often expected to have no effect and are just included in experiments to avoid a potential threat to validity.  As discussed above, it can make sense to include a sham treatment in the design even if its effects might be small, but then the analysis should allow for that possibility, which in our model is done by including the mean and variance of the sham effects as hyperparameters which will be near zero in that case. 

\section{Applied example 2:  Meta-analysis of sham-controlled trials of rTMS for treating major depression}\label{berlim}

\subsection{Background}
Depression is a highly prevalent public health issue with enormous social and economic cost.
For a large group (20--30\%) of patients, existing treatments do not suffice to achieve remission.
Moreover, existing treatments can take a long time to achieve remission if they do at all and tend to be
associated with unpleasant side effects.
For this reason, new treatment options for major depression are badly needed.
In the last two decades, repeated transcranial magnetic stimulation (rTMS) has
emerged as a promising, non-invasive new treatment option.
Specifically, rTMS treatment is achieved by inducing electric currents within the 
brain by applying a changing magnetic field (generated by electricity running through 
a coil of wire near the scalp of the patient).

Since its introduction as a potential treatment for depression, rTMS has been studied
in a large number of randomized control trials.
Recently, Berlim et al.\cite{Berlim} published a highly cited systematic review and
meta-analysis of such trials to assess the suitability of 
rTMS as a treatment for major depression, estimating the response, remission and
dropout rates.
This meta-analysis included 29 suitable randomized, double-blind, sham-controlled trials out of
the 396 such trials they previously identified; in the present paper, we analyze the 15 of these trials that include data on remission rates.
In each trial, patients were exposed to a real or sham rTMS treatment, 
consisting of a coil angled on the scalp or the use of a specific sham coil.

\begin{table}
  \begin{small}
    \centerline{
      \begin{tabular}{c cc cc}
                       & \multicolumn{2}{c}{Sham treatment} &
        \multicolumn{2}{c}{Real exposure} \\
        Study name & Remission $n_{j0}$ & Total $N_{j0}$
        & Remission  $n_{j1}$ & Total  $N_{j1}$ \\
        \hline
        \ George et al.\ (1997)            & 0 & 5 & 1 & 7 \\
        \ Berman et al.\ (2000)            & 0 & 10 & 1 & 10 \\
        \ \vdots &\vdots & \vdots &\vdots &\vdots \\ 
        \ Bakim et al.\ (in press)  & 1 & 12 & 9 & 23
      \end{tabular}
    }
  \end{small}
  \caption{\em A portion of the data used for the rTMS meta-analysis
    reported by Berlim et al.\cite{Berlim} \ The data consist of remission and total
    counts observed in all the included studies for both real and sham rTMS treatment.}
  \label{rawdataberlim}
\end{table}

Berlim et al.\cite{Berlim} report odds ratios between the real and sham treatments for response, remission, and dropout rates, for both the individual studies included in the meta-analysis as well as for
the compound meta-analysis.
In the present paper, we focus on an alternative analysis of the remission rates, but we could have equally
well have chosen the response or dropout rates.

A subset of the data used by Berlim et al.\cite{Berlim} are displayed for clarity in Table \ref{rawdataberlim}.
For study $j$, these data are remission counts $n_{j0}$ out of a total of $N_{j0}$ patients for the sham treatment as well as remission counts $n_{j1}$ out of a total of $N_{j1}$ patients for the real treatment.

\subsection{Originally published analysis}\label{berlim-orig}

\begin{figure}
  \centerline{
    \includegraphics[width=.5\textwidth]{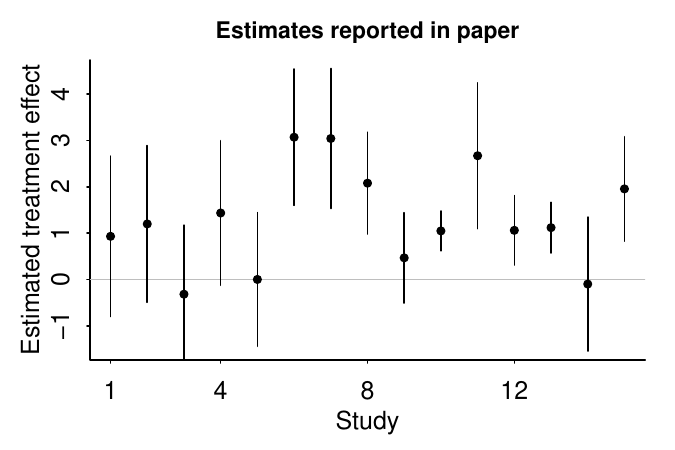}
    \includegraphics[width=.5\textwidth]{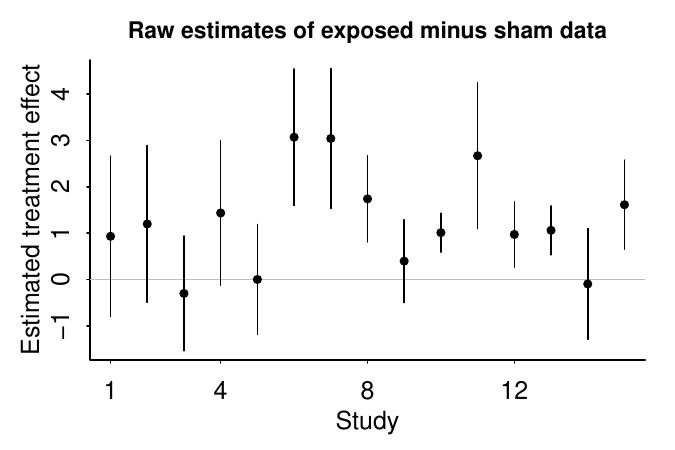}
  }
  \vspace{-.1in}
  \caption{\em (a) Estimated effects for each of the rTMS experiments as reported by Berlim et al.\cite{Berlim} \
  (b) Raw difference estimates $y_{j1}-y_{j0}$ with standard errors $(s_{j1}^2+s_{j0}^2)^{1/2}$.
  }
  \label{berlim-raw}
\end{figure}

From this data, Berlim et al.\cite{Berlim} calculate estimates and confidence intervals of the odds ratio
of the two treatments, using the hierarchical modeling approach of DerSimonian and Laird\cite{DerSimonianLaird1986}.
These estimates can be understood in our framework as follows.

We can straightforwardly calculate the log odds of remission $y_{j0}=\log((n_{j0}+0.5)/(N_{j0}+1))$ for the sham experiments and $y_{j1}=\log((n_{j1}+0.5)/(N_{j1}+1))$ for the real exposures. As is commonly done, we
deal with cells with zero counts by adding a Haldane-Anscombe correction of 0.5.
Assuming a binomial distribution of the remission counts, the log odds will be approximately normally distributed
for large enough sample sizes.
We can apply the power method to derive estimates $s_{j0}$ and $s_{j1}$ for the standard errors of $y_{j0}$ and $y_{j1}$, respectively.
To be precise, we estimate $s_{ji}=\sqrt{(n_{ji}+0.5)^{-1}+(N_{ji}-n_{ji}+0.5)^{-1}}$.
We are now again in a position where we can think of $y_{ji}$ arising from  model (\ref{2equations}).

Their estimates of the log odds ratio are reproduced in Figure \ref{berlim-raw}a.
They can be seen to be compatible with the plain difference estimates in Figure \ref{berlim-raw}b.
Both the estimates and standard errors corresponding to different experiments
vary widely across studies (particularly considering that these are log odds).

\subsection{Exploration of the sham data}

\begin{figure}
  \centerline{
    \includegraphics[width=.5\textwidth]{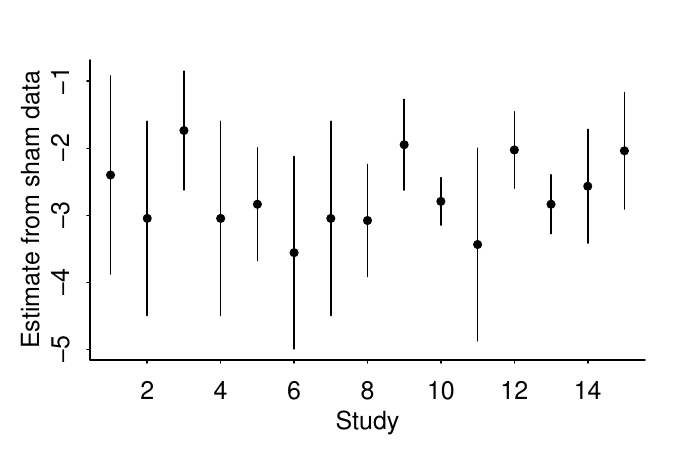}
    \includegraphics[width=.5\textwidth]{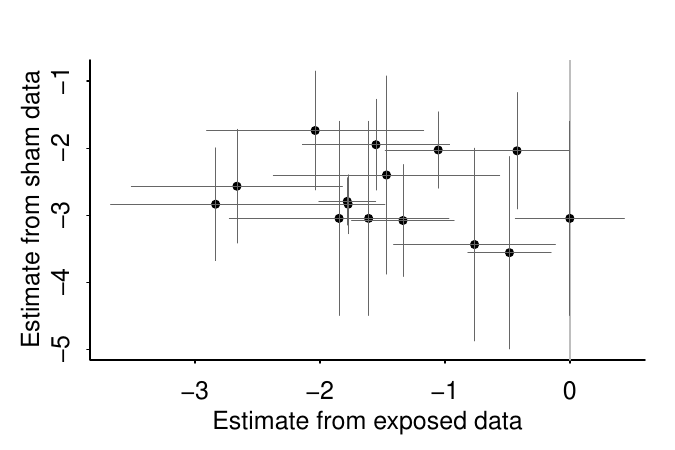}
  }
  \vspace{-.1in}
  \caption{\em (a) Estimates $\pm$ standard errors of the effect of the sham
    treatments for the studies in the meta-analysis of rTMS.
    (b) Sham vs.\ exposed estimates.}
  \label{berlim-sham}
\end{figure}

The nature of the sham data in this second example is different than in the example of Section \ref{chicks}.
Indeed, it would make not any sense for our sham data to be noise (centered at zero), as log odds
of zero would correspond to even odds, that is, a coin flip.
But a treatment for major depression with a remission rate of 50\% would constitute a major breakthrough.
Therefore, we would expect our sham data to look like anything but noise.
Indeed, upon inspection, this turns out to be the case, as shown in Figure \ref{berlim-sham}.

This immediately makes clear that discarding the sham data and working with the exposed data alone
is not an option.
However, our hierarchical model (\ref{meas}) still can provide a superior alternative as it allows us to reconsider our estimates of the sham and treatment effects in the individual studies in the light of the larger meta-analysis, by pooling them toward their common mean.

\subsection{Reanalysis using the multilevel model}\label{berlim-posterior.fit}

\begin{table}
  \begin{small}
    \centerline{
      \begin{tabular}{crr}
        Parameter          & Estimate (s.e.) & 95\% interval        \\\hline
        $\mu^{\theta}$     & 1.2 (0.3)   & $\ \ [0.7, 1.8]$ \\
        $\sigma^{\theta} $ & 0.5 (0.3)   & $\ \ [0.0, 1.1]$ \\
        $\mu^{b}$          & $-$2.5 (0.2)   & $[-2.9, -2.1]$    \\
        $\sigma^b$         & 0.3 (0.2)   & $\ \ [0.0, 0.7]$ \\
      \end{tabular}
    }
  \end{small}
  \caption{\em Posterior means, standard deviations, and 95\% intervals  for the
    hyperparameters in the hierarchical model fit to the rTMS
    data.}\label{berlim-inferences}
\end{table}

We now fit the multilevel model (\ref{meas}) to the Berlim et al.\ data; inferences for the
hyperparameters appear in Table \ref{berlim-inferences}.  The estimates of
$\mu^{\theta}$ and  $\sigma^{\theta}$ imply a distribution of treatment effects
with a clearly positive mean, along with substantial variation, implying
different effects across different studies. We can interpret the estimated treatment 
effect $\mu^{\theta}$ of $1.2$ as saying that the odds of remission when receiving the real treatment
are about three and a half ($\exp(1.2)$) times as good as when being treated with the sham.
We can interpret our estimated sham effect $\mu^{b}$ of $-2.5$ as saying that even with the 
sham treatment, there is still a probability of about $1/13$ of remission.

Figure \ref{hierarchical}a shows the posterior mean and standard deviation of
the treatment effect $\theta_j$ for each experiment $j$.  For comparison, we
display in Figure \ref{hierarchical}b the raw difference estimates $y_{j1}-y_{j0}\pm(s_{j1}^2+s_{j0}^2)^{1/2}$.
The estimates from the hierarchical model have been
partially pooled toward the common mean but otherwise show a pattern similar to
that of the difference estimates, with the largest changes being the 
studies that with extreme conclusions and large standard error. These were thus pulled closer to the center of the
distribution.

\begin{figure}
  \centerline{
    \includegraphics[width=.5\textwidth]{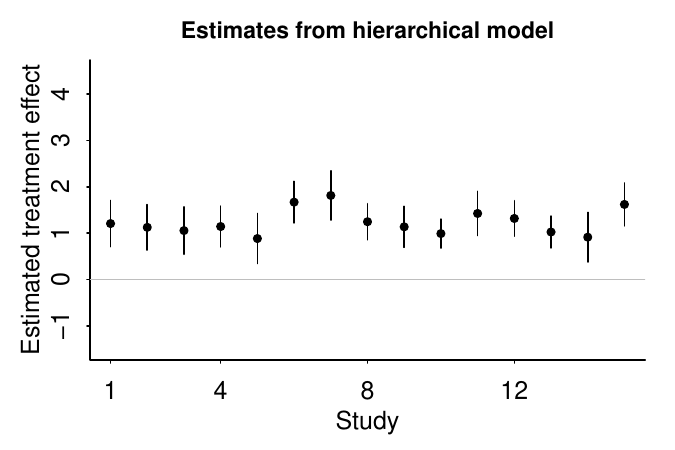}
    \includegraphics[width=.5\textwidth]{berlim4h.pdf}}
  \vspace{-.1in}
  \caption{\em  (a) Posterior mean $\pm$ standard deviation of each treatment
    effect $\theta_j$ from the hierarchical model fit to the rTMS data. The
    fitted model did not estimate the sham effects to be zero, and so these estimated treatment
    effects take into account both the exposed and sham data.
    (b) For comparison, the raw difference estimates $y_{j1}-y_{j0}$ $\pm$ standard error.
    The Bayesian hierarchical estimates on the left
    plot are partially pooled toward a common mean.}
  \label{hierarchical}
\end{figure}

Seeing that we fit our models in Stan, there is nothing forcing us to make a normal
approximation to the likelihood and we could in fact have worked just as easily with
a binomial observation model.
This does not substantively alter the conclusions, though it ends up pooling the studies slightly less and leads to larger estimates of uncertainty.
We discuss this in Appendix \ref{alt-mod-binomial}.

\section{Evaluating the competing estimates using simulation}\label{evaluation}
We have seen the hierarchical model work on two real problems, one where there was no
evidence of sham effects and one where correction for the sham was necessary.  We can better understand how the model works by using simulation to set up and evaluate a series of scenarios with different levels of strength of the sham signal.  We demonstrate this approach by perturbing the chick brain example of Section \ref{chicks}.

\subsection{Setting up a family of hypothetical scenarios} \label{subsec:simulation}
To see what happens when different levels of sham correction
is necessary, we study a series of simulated examples indexed by a parameter
tied to the size of the sham effects.  We can then compare the three
estimates---(a) the exposed data estimate, $y_{j1}$, (b) the difference between
exposed and sham, $y_{j1}-y_{j0}$, and (c) the hierarchical model estimate
$\mbox{E}(\theta_j|y)$---and see how they perform as a function of the scale of
the bias parameters, $b_j$.

We set $\mu^b$ to 0 and consider a range of values for $\sigma^b$, for each
performing the following steps 200 times:  (1) Simulate one draw of the vector
of 38 values $b_j, j=1,\dots,J$, drawing them independently from the
$\mbox{normal}(0,\sigma^b)$ distribution; (2) Draw the vector of the 38 values
$\theta_j,j=1,\dots,J$, from their (joint) posterior distribution\footnote{In Appendix
\ref{simulation-alt}, we show that the conclusions of our simulation study remain 
the same if we work with the raw estimates of $\theta_j$ instead.} from Section
\ref{posterior.fit}; (3) Simulate one dataset, that is a vector of 38 values
$y_{j0}\sim \mbox{normal}(b_j, \sigma^y)$ and a vector of 38 values $y_{j1}\sim
\mbox{normal} (\theta_j+b_j, \sigma^y)$.

We are assuming that the residual scales $\sigma^y$ are known and all equal to
0.04, a value chosen because it is approximately the average of the standard
errors in the data; see Table \ref{rawdata}.  This simplification, along with
that of assuming $\mu^b=0$, makes it easier to interpret the results of our
simulation but should not materially affect our results.  We explore the role of
$\mu^b$ in Section \ref{linear.approx}.

\subsection{Evaluations}

\begin{figure}
  \centerline{\includegraphics[width=\textwidth]{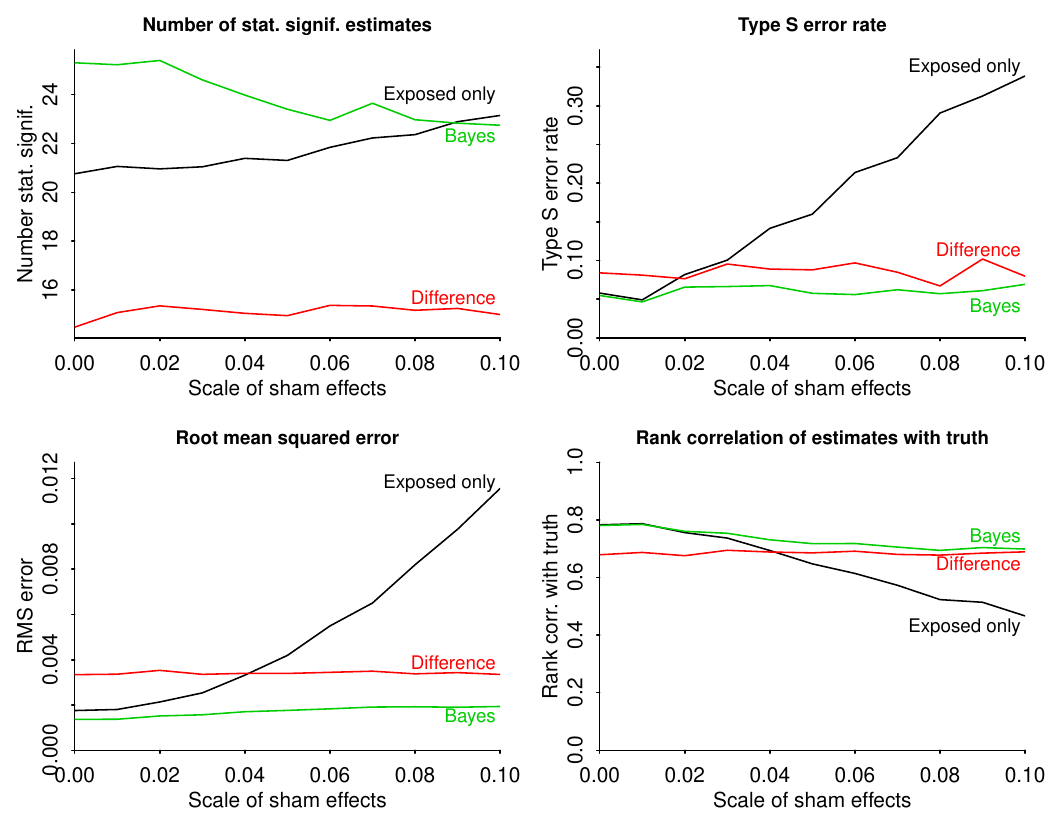}}
  \vspace{-.1in}
  \caption{\em Results of simulation study comparing three estimates--- the
  exposed data estimate, $y_{j1}$, the difference between exposed and sham,
  $y_{j1}-y_{j0}$, and the Bayesian hierarchical model estimate
  $\mbox{E}(\theta_j|y)$---to simulated data.  The four graphs show the results
  for four different frequency evaluations, and on each graph the horizontal
  axis represents $\sigma^b$, the standard deviation of the sham effects in the
  simulation.}
  \label{simulation}
\end{figure}

As discussed in Section \ref{freq}, for each set of simulated parameters and
data, we then compute the following four summaries for each of the three
estimates:  the proportion of the 38 estimates that are statistically
significant, the type S error rate, the mean squared error of the 38 estimates, and
the rank correlation between the estimates and the 
true $\theta_j$'s.

We choose a grid of values of $\sigma^b$ between 0 and 0.10, choosing that upper
bound as this is the approximate standard deviation of treatment effects (see
inference for $\sigma^{\theta}$ in Table \ref{inferences}), and we would not
expect the variation in sham effects to be higher than the variation in
treatment effects.

For each  $\sigma^b$ we average each of the above summaries over our 200
simulations to obtain $3\times 4$ matrix of four frequency evaluations for the
three estimates:  exposed, exposed minus sham, and hierarchical Bayes.
Figure \ref{simulation} plots the results for each frequency evaluation as a
function of $\sigma^b$, the scale of the sham effects.  The difference estimate,
$y_{j1}-y_{j0}$, outperforms the exposed-only estimate $y_{j1}$ when sham
effects are large (the right side of each graph) but not when sham effects are
small.  The Bayesian hierarchical estimate outperforms both, in part by appropriately
managing the sham data and in part by pooling across experiments.

We now go through each of the frequency properties for this example:
\vspace{-3pt}
\begin{itemize}
  \item {\em Number of statistically significant claims:}  The difference
        estimate yields the lowest rate of statistically significant results,
        which makes sense given that it is the noisiest of the estimates.  When
        sham effects become large, the rate of apparently statistically
        significant estimates from the exposed data alone goes up, but this is
        an illusion based on the fact that the variance in the data is
        increasing but this is not reflected in the standard errors.
      \item {\em Type S error rate:}  The difference and Bayes estimates have
        approximate 5\% type S error rates, as does the exposed-only estimate
        when the sham effects are negligible.  As sham effects become larger,
        the error rate for the exposed-only estimate becomes increasingly
        unacceptable.
      \item {\em Root mean squared error:}  The Bayes estimate performs the best,
        unsurprisingly as it makes use of the most information, and we are
        simulating from the model.  The exposed-only estimate outperforms the
        difference estimate when sham effects are near zero---this is what we
        saw in Section \ref{chicks}---but when sham effects are large, the
        exposed-only estimate has a huge error.
  \item {\em Rank correlation with truth:}  When sham effects are small, the
        exposed-only estimate is best; when sham effects are large, the
        difference is best; in all cases the Bayes performs as well
        as the other two. At each extreme, the Bayes does as well as, not better
        than the corresponding simple estimate; this is because, in this simple
        simulation where the error variances for all experiments are equal, the
        partial pooling across experiments affects estimates and standard errors
        but does not alter the ranking of the 38 estimates.
      \end{itemize}
              \vspace{-3pt}
The results shown in Figure \ref{simulation} are consistent with the idea of the
difference being a conservative estimate---and, indeed, {\em if} the only
available choices were the exposed-only and the difference estimate, {\em and}
no information were available regarding $\sigma^b$, the scale of the sham
effects, then we might well prefer the difference as the safe option. In fact,
though, we are also free to use the hierarchical Bayes estimate, and even if
that were not available, the data are informative about $\sigma^b$, so we would
not recommend the difference estimate as a default analysis.

\section{Linear adjustment via partial Bayesian inference}\label{linear.approx}

We can gain intuition about the sham-adjustment problem by considering a partially
Bayesian model in which the sham effects come from a
$\mbox{normal}(\mu^b,\sigma^b)$ distribution but the treatment effects
$\theta_j$ are estimated using maximum likelihood (equivalently, Bayesian
inference with $\sigma^{\theta}$ set to infinity and $\mu^{\theta}$ becoming
irrelevant). This can also be viewed as a measurement error model, where the
$y_{j0}$'s are noisy measurements of latent variables $b_j$. To simplify the
algebra, we assume a normal likelihood for the measurements.

Under any of these formulations, fixing the hyperparameters results in linear
estimates for the $\theta_j$'s, which in turn allows clear comparisons with the
exposed-only and difference estimates.  This is related to the work of Turner et al.\cite{TurnerJacksonWeiThompsonHiggins2015} to make Bayesian meta-analysis more accessible using analytic formulas that approximate fully Bayesian inferences.

To work out the solution algebraically it is convenient to first perform
inference for the sham effects. Combining the prior distribution, $b_j\sim
\mbox{normal}(\mu^b,\sigma^b)$, with the sham measurement,
$y_{j0}\sim\mbox{normal}(b_j, \sigma_{j0})$, yields a posterior distribution,
$b_j\sim\mbox{normal}(\hat{b}_j, s_j)$, where
$$\hat{b}_j= \frac{\frac{1}{(\sigma^b)^2}\mu^b +
\frac{1}{\sigma_{j0}^2}y_{j0}}{\frac{1}{(\sigma^b)^2} +
\frac{1}{\sigma_{j0}^2}}\qquad \textnormal{and}\qquad s_j=
\left(\frac{1}{(\sigma^b)^2}+ \frac{1}{\sigma_{j0}^2}\right)^{-1/2}.$$ The
corresponding maximum likelihood estimate  $\hat{\theta}_j$ is  $y_{j1}
-\hat{b}_j$, which can be written as
\begin{equation}
  \hat{\theta}_j= y_{j1} - \mu^b - \lambda (y_{j0} - \mu^b)
  \label{simplified}
\end{equation}
with standard error $\sqrt{s_j^2 +\sigma_{j1}^2}$, and where
$$
  \lambda = \frac{(\sigma^b)^2}{(\sigma^b)^2+ \sigma_{j0}^2}
$$
is the variance ratio which determines the amount by which the exposed-data
estimate must be adjusted for the sham measurement.

The estimate  (\ref{simplified}) reduces to the exposed-only estimate when
$\mu^b=\sigma^b=0$ (that is, when there are no sham effects) and reduces to the
difference estimate as $\sigma^b\rightarrow\infty$ (as sham effects become
large).  The standard error  of $\hat{\theta}_j$ reduces to $\sigma_{j1}$ when
sham effects are zero and $\sqrt{\sigma_{j0}^2 + \sigma_{j1}^2}$ in the limit of
large sham effects.

In between these extremes, equation (\ref{simplified})---the maximum likelihood
estimate under the measurement error model---is constructed by first subtracting
the average sham effect, which represents the average bias for all the
experiments---and then a subtracting a fraction of the relative estimated sham
effect from experiment $j$, with that fraction depending on the relative values
of $\sigma^b$ and $\sigma_{j0}$.  For the chicken data, $\mu^b$ is estimated to
be essentially zero and $\sigma^b$ is estimated to be much smaller than
$\sigma_{j0}$ for all the experiments (see Table \ref{inferences}), so there
is no essentially no need to adjust for the sham measurements.

In practice we would recommend full Bayesian inference as in Section
\ref{bayes}.  Or, if there is reluctance to partially pool across experiments,
one could fit the same Bayesian model but removing the prior on the $\theta_j$'s
(equivalently, constraining $\sigma^{\theta}$ to $\infty$; see Appendix \ref{appx:removing-partial-pooling}). The point of the
above algebra is just to clarify the way in which the optimal estimate of
treatment effects will in general approximately take the observed estimate $y_{j1}$ and
subtract some fraction, between 0 and 1, of the sham estimate.

\section{Discussion}\label{discussion}

\subsection{Failure modes and limitations of the method}
For the reasons discussed above, we prefer the hierarchical Bayesian model to
the simple treated-only or difference analyses:  we think that the estimates
obtained from our model are more reasonable and that they would yield better
predictions in a replication study.  But there must be settings where our
approach would perform poorly.  When will that occur?

Speaking generally, Bayesian inference with noisy data works by partial pooling
toward a fitted model.  When the fitted model is wrong, the pooling can go in
the wrong direction, yielding poor inferences.  In the problem discussed in this
paper, the sham and treatment estimates are each pooled toward the mean of that
set of experiments.  For the sham, this does not seem to be a problem, first
because we expect sham effects to be small, second because we have no reason to
expect patterns in the sham effects.  If we did expect such patterns, it would
make sense to include them in the model, for example by allowing a correlation
between sham and treatment effects as discussed in Appendix \ref{alt-mod-corr}.
For the treatment effects, partial pooling toward a common mean could be more of
a concern, for example if there is a trend or if the pattern of effects is
otherwise predictable.  This is related to the problem of edge effects when
estimating a function from noisy data:  an extrapolative model can overfit
trends in the data, but a model that is more conservative in its extrapolation
can flatten out at the edges.  Ultimately one must accept that inferences are
sensitive to uncheckable assumptions.

For our default hierarchical model to fail badly, two things must happen: 
the data must be noisy enough for the partial pooling to make a difference,
and the underlying trend or pattern must itself be strong. Both these things
can happen, for example if the treatment effects follow a linear trend. In our two applied
examples, there was no apparent trend in the data; had there been, it
surely would have been included in the model.  But a proposed statistical method
will be used in all sorts of settings.  Were we to fit our no-trend hierarchical
model to data with an actual trend, we would overestimate effects at the low end
and underestimate at the high end, in aggregate understating the
variation in the treatment effects. In this case, our recommended solution would
be to incorporate this possible trend by adding it into the mean of the distribution
for $\theta_j$ in (\ref{meas}).

More generally, nonlinear models are possible, hence it can make sense to check
sensitivity of analyses to various choices of model, as we demonstrate in
Appendix \ref{alt-mod}.  We prefer our default hierarchical model to the simple
default of exposed minus sham, but in general it makes sense to consider
scientifically plausible alternatives as well.  This is an unavoidable concern
when using measurement error or latent variable models, but ultimately we see no
good alternative to modeling, as the simple unpooled estimates are just too
noisy and wasteful of data.

The main practical limitation of our method is that it works best when there is a large number of repeated studies.  We explore the number of studies needed to benefit from our method in Appendix 
\ref{simulation-sizes}. When the number of studies is small, the hierarchical model can still be fit, but the user would be advised to include strong prior information on the hyperparameters. This could well be a good idea---indeed, we would prefer it to a riskier or noisier strategy such as fully subtracting or ignoring the sham data---but we recognize that a fully Bayesian approach would put more of a burden on many researchers. Hence in the present paper we focus our recommendations on the problem of repeated studies.

\subsection{Chick brains experiment}

A key point of this paper is that our analysis could have made a difference in our motivating example, both in the conclusions drawn from the existing data and in the design of the study.

Blackman\cite{Blackman2015} wrote that his team ``worked very closely with a statistician$\,$\dots to optimize our procedures for maximum statistical power.'' 
Care went into both the scientific and statistical aspects
of the design of the study, as well as the data collection itself. This is one
indication of the potential importance of the statistical modeling and analysis
plan we have presented here:  if a team of conscientious researchers, working on
a policy-relevant research program and aware of cost constraints and the
importance of statistical efficiency, can perform an analysis that is
mathematically equivalent to discarding half the information in their data, this
represents large gains from a new paradigm, moving away from cookbook rules to
an open-ended modeling approach.  Indeed, Blackman\cite{Blackman2015} also writes, ``Plans
were made to follow up \dots but the experiment could not be brought to
fruition.''  In this case, discarding the sham data would have been equivalent to
doubling the sample size of the experiments, without any additional data-collection cost at
all.

% In retrospect it would have been enough to collect a smaller set of sham data in
% the chick brains study; there was no need to replicate all 38 experiments. This
% was not clear a priori but is apparent upon examination of the sham data. A more
% efficient, sequential, design would recommend gathering {\em some} sham data,
% but once its irrelevance becomes clear (as seen from the estimated values of
% $\mu^b$ and $\sigma^b$ in the hierarchical model), not so much would need to be collected.
% Given the uncertainty in some of the frequency comparisons of interest (an
% uncertainty masked by the illusion of informativeness of comparisons of
% $p$-values), limited experimental resources could have been used more effectively
% by collecting more data on non-sham treatments.  We do not consider this as a
% devastating criticism of the study---it is unfortunately all too common,
% including in our own work, to gather data in rectangular structure with an eye
% toward convenience rather than efficiency---but it is worth considering these
% issues when designing future experiments.

\subsection{More general implications for design and analysis of structured
  experiments}

What is striking about the results from this paper, as distinguished from many
other examples of the practical efficiency gains that can be obtained from
Bayesian inference,
is how simple and effectively the Bayesian approach works out in this example,
requiring no specialized knowledge or custom prior distributions.  This gives us
hope that hierarchical modeling can resolve other common data-combination
problems in applied statistics, and it is why we have been continuing to chew on
this example for thirty years.

Specifically, we recommend our Bayesian multilevel model as a default analysis for repeated controlled experiments.
Indeed, it gives more efficient estimates than both the commonly used difference
or exposed-only estimates. More importantly still, it systematically determines
from the data how much adjustment for the sham measurements is appropriate, by
interpolating between the extremes of difference and exposed-only estimates,
rather than leaving that choice to the scientist.

There is some awkwardness that, in order to perform our more efficient estimate, we need to model the treatment effects $\theta_j$ and the biases $b_j$.  This is a general property of measurement-error models, and we believe there are many cases, including the examples described in the paper, where the small effort in constructing probability models for treatment effects and biases is minor compared to the efficiency gains obtainable from the model-based estimate.  An alternative for those who would prefer not to
partially pool the $\theta_j$'s across experiments is to use the procedure of Section \ref{linear.approx} to find an optimal linear adjustment, which is corresponds to modeling just the biases without partially pooling the treatment effects.

% On top of that, we suggest a sequential experimental design, in case of costly
% sham data collection. If, in the course of data collection, the recommended
% analysis confidently estimates the sham effects to be substantively
% insignificant, collection of sham data can be halted and the resources can be
% transferred, for instance, to collection of more exposed data.

Combining these two recommendations for the statistical analysis and
experimental design of controlled experiments should enable a more
cost-effective scientific practice. We hope this will contribute to an
increase in replicable scientific findings.

\noindent

\bibliographystyle{ama}
\bibliography{references}

\clearpage
\appendix

\section{Alternative models for example 1}\label{alt-mod}
In this appendix, we discuss some alternative models we could have used for
analyzing the chick brain data from Section \ref{chicks}.

\subsection{Measurement error with correlation}\label{alt-mod-corr}

\begin{figure}
  \centerline{
    \includegraphics[width=.5\textwidth]{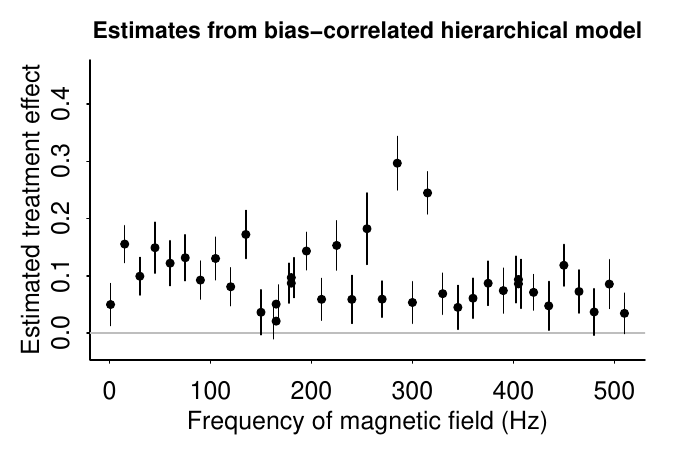}
    \includegraphics[width=.5\textwidth]{blackman4a.pdf}}
  \vspace{-.1in}
  \caption{\em  (a) Posterior mean $\pm$ standard deviation of each treatment
    effect $\theta_j$ from the hierarchical model with correlated bias and
    treatment effects fit to the chick data. (b) For comparison, the estimates
    of the original hierarchical model without correlations, which can be seen
    to be almost identical.}
  \label{correlated}
\end{figure}

It is conventional with measurement error models to use independent errors, and
this is what we did in (\ref{meas}), with the idea being that there can be an
average sham effect and variation in the sham effects but with no correlation
expected with the treatment effects.  This makes sense in the chick experiment,
as the treatment effect varies by frequency of the magnetic field, whereas the
bias or sham effect should have nothing to do with frequency.

More generally, though, one might want to allow the treatment effect and its
measurement bias to be correlated, in which case (\ref{meas}) can be generalized
to a bivariate normal distribution for $(\theta_j,b_j)$ with a covariance
matrix.  Figure
\ref{correlated} shows the results of fitting this to the chick data; these treatment effect estimates are essentially the same as from the uncorrelated-errors model fit in Section \ref{posterior.fit}.

\subsection{Gaussian process for the treatment effects}

\begin{figure}[h]
  \centerline{
    \includegraphics[width=.5\textwidth]{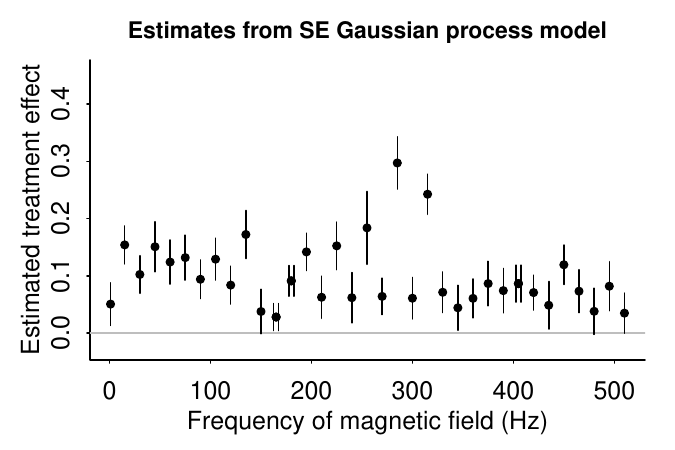}
    \includegraphics[width=.5\textwidth]{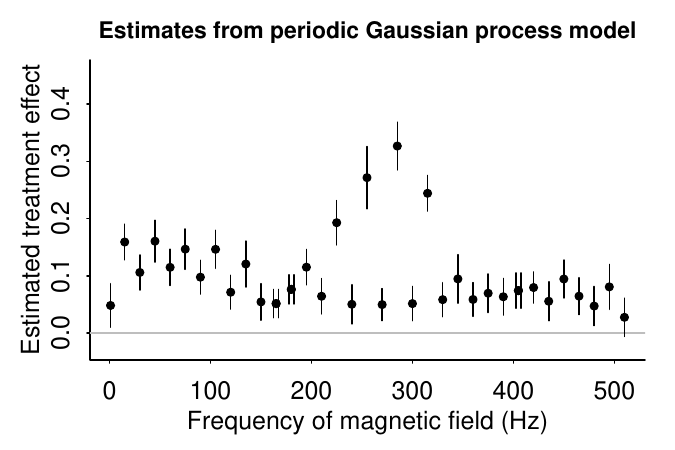}}
  \vspace{.1in}
  \centerline{
    \includegraphics[width=.5\textwidth]{blackman4a.pdf}
    \includegraphics[width=.5\textwidth]{blackman4b.pdf}}
  \vspace{-.1in}
  \caption{\em  {\em Top row:}  (a) Posterior mean $\pm$ standard deviation of
    each treatment effect $\theta_j$ from the squared-exponential  (SE) kernel
    Gaussian process model fit to the chick data. (b) The corresponding plot for
    the periodic kernel Gaussian process model. The fitted models estimate the
    sham effects to be essentially zero, and so these estimated treatment
    effects come pretty much from the exposed data alone. The Bayesian estimates
    in the left plot are partially pooled toward each other for close
    frequencies. The estimates in the right plot are partially pooled toward
    each other for frequencies whose difference is close to 30 Hz.\newline
    {\em Bottom row:} For comparison, (c) the estimates from our default
    analysis and (d) the raw estimates $y_{j1}$ from the exposed data.}
  \label{gaussian-process}
\end{figure}

A potential concern regarding the models fit so far is that they do not encode
any structure in the treatment effects.  One challenge here is that so many
different structures are possible, as discussed in the original Blackman et al.\cite{BlackmanBenaneElliottHousePollock1988} paper.  As discussed in Section \ref{orig}, various complicated patterns
of alternating frequencies were extracted, but many of these conclusions are
shaky as they rely on inherently noisy comparisons of $p$-values.

We think the measurement error model of Section \ref{bayes} is a sensible
default analysis, but more structured models would be possible.  As an example,
we consider two Gaussian process (GP) models for the vector $\theta$ as a
function of frequency:  one model which favors local smoothness of the treatment
effects (a GP with a squared-exponential covariance function) and
one which favors similar effects for frequencies separated by 30 Hz 
(a GP with a periodic covariance function with a period of around 30 Hz).

The resulting estimates are displayed in Figure \ref{gaussian-process}. The squared-exponential GP model gives estimates that are very close to
those of the hierarchical model. This model favors stronger pooling between
measurements which are close in frequency. This results, for example, in a
slightly higher estimate for the treatment effect at 285 Hz but it is most
visible for the frequencies with repeated measurements which it forces to have
the same estimated effect sizes. For the periodic GP model, we observe
interestingly different estimates compared to our default analysis, due to the
periodic partial pooling behaviour it enforces. For example, we see that the
estimates at 225 and 345 Hz are pulled upwards, a phenomenon we do not observe
in our default analysis.

One difficulty in using such GP models for analyzing the data is the question of
how to choose an appropriate prior on the length-scale parameter. This parameter
regulates the scale on which the smoothing happens. That is to say, it
determines how close two frequencies need to be to each other in order to
qualify to be pooled together. This prior should be chosen based on domain
expertise in each particular application. We believe this makes the GP analyses
less suitable as a default choice, unless strong domain knowledge of that kind
is available.

\subsection{Removing partial pooling}\label{appx:removing-partial-pooling}

\begin{figure}
  \centerline{
    \includegraphics[width=.5\textwidth]{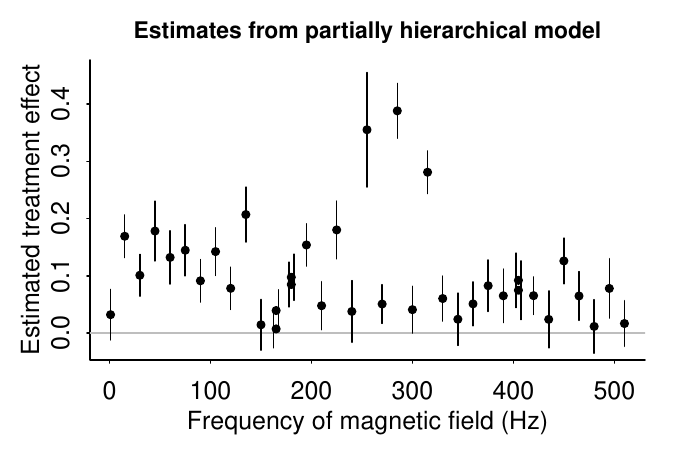}
    \includegraphics[width=.5\textwidth]{blackman4b.pdf}}
  \vspace{-.1in}
  \caption{\em  (a) Posterior mean $\pm$ standard deviation of each treatment
    effect $\theta_j$ from the partially-hierarchical model, which partially
    pools the biases but not the treatment effects, as fit to the chick data.
    The fitted model estimated the sham effects to be essentially zero (see
    discussion of Table \ref{inferences}), and so these estimated treatment
    effects come pretty much from the exposed data alone. (b) For comparison,
    the raw estimates $y_{j1}$ from the exposed data. These two
    estimates roughly coincide.}
  \label{partially-hierarchical}
\end{figure}

\begin{figure}
  \centerline{
    \includegraphics[width=.5\textwidth]{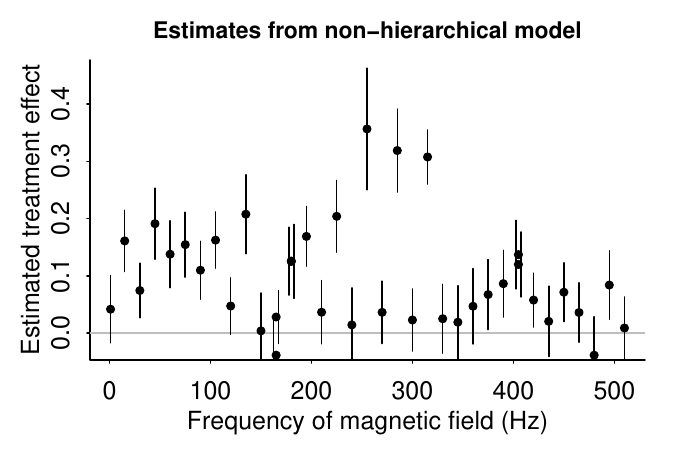}
    \includegraphics[width=.5\textwidth]{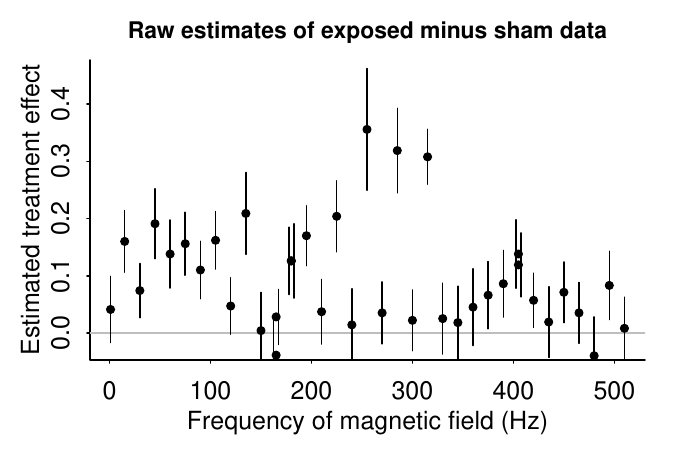}}
  \vspace{-.1in}
  \caption{\em (a) Posterior mean $\pm$ standard deviation of each treatment
  effect $\theta_j$ from the non-hierarchical model, which partially pools
  neither the biases nor the treatment effects, as fit to the chick data. (b)
  For comparison, the raw difference estimate $y_{j1}-y_{j0}$ for the chick data
  $\pm$ standard error. These two estimates roughly coincide.}
  \label{non-hierarchical}
\end{figure}

Following Section \ref{linear.approx}, it may be interesting to inspect the
estimates given by variants of the hierarchical model, where we first remove the
partial pooling of the treatment effects and next also that of the biases.

When we remove the partial pooling of the treatment effects (equivalent to the limit,
$\sigma^\mu\to\infty$), but keep partial pooling of the biases, we estimate that
 $\mu^b$ and $\sigma^b$ are both near zero, as would be expected from Table
\ref{rawdata}. As anticipated by the algebra of Section \ref{linear.approx}, we
obtain, in effect, the raw exposed-only estimates. This is shown in Figure
\ref{partially-hierarchical}.

When we additionally remove the partial pooling of the biases (equivalent to the limit,
$\sigma^b\to\infty$), the algebra of Section \ref{linear.approx} would predict
that we roughly end up giving the raw difference estimate. Indeed, we see this
confirmed in Figure \ref{non-hierarchical}. The difference estimate is similar to the exposed-only estimate but is much higher in
uncertainty. Dropping the partial pooling of the biases has the same result of
increasing the noise in our estimates.

The two estimates of Figure \ref{non-hierarchical} always
coincide, but the collapse of the two estimates of Figure
\ref{partially-hierarchical} only happens when the sham data is effectively
noise. The partially hierarchical model that partially pools the biases but not
the treatment effects might be a superior alternative to the exposed-only and
difference estimates in case there is reluctance to partially pool across
experiments as we do in our default analysis.  

\subsection{A Bayesian meta-analysis of the difference estimates}\label{bayesian-metaanalysis}
A different conceivable analysis (which tends not to be used in practice, as far as we are aware)
for our examples is to perform a standard Bayesian meta-analysis, as described for example 
in Gelman et al.\cite{GelmanCarlinSternDunsonVehtariRubin2013}, section 5.6, on the difference estimates.
This would amount to the model ,
\[
  \begin{array}{rl}
y_{j1}-y_{j0} &\sim \mathrm{normal}(\theta_j, \sqrt{s_{j1}^2+s_{j0}^2})\\
\theta_j &\sim \mathrm{normal}(\mu^{\theta},\sigma^{\theta}).
  \end{array}
\]
Where our proposed default analysis separately models the treatment and 
sham effects, similar to a measurement error model, this analysis is a standard 
hierarchical model fit to the difference estimates.
Figure \ref{bayesian-metaanalysis} shows that this simplified 
hierarchical model is wasteful of important information 
in the data, compared to our default hierarchical analysis that models the sham and exposed data jointly.

For example, consider the frequency 255 Hz.
The raw difference estimates (see Figure \ref{non-hierarchical}b) and the estimates from the 
Bayesian meta-analysis of these difference estimates (Figure \ref{bayesian-metaanalysis}b) agree that 
this is expected to be the frequency with the largest treatment effect.
However, at this frequency, the estimate from exposed data has large uncertainty compared to that from the sham data.
As a consequence, an analysis that makes better use of the available data would substantially correct 
the estimates for this frequency downwards. This is precisely what our proposed default analysis achieves, by 
separately modeling the exposed and sham effects (see Figure \ref{bayesian-metaanalysis}a).
Our default analysis estimates largest treatment effects to be at 285 Hz, with the estimate from this frequency being corrected downwards less, because its
exposed data reveal less uncertainty compared to the sham data.
An analysis working only with the differences between the sham and exposed data would discard 
important information such as the respective uncertainty in the sham and exposed measurements. 
It is for this reason that we prefer to model both the sham and exposed data explicitly and it is why 
we do not recommend the Bayesian meta-analysis of the differences as our default analysis.

\begin{figure}
  \centerline{
    \includegraphics[width=.5\textwidth]{blackman4a.pdf}
    \includegraphics[width=.5\textwidth]{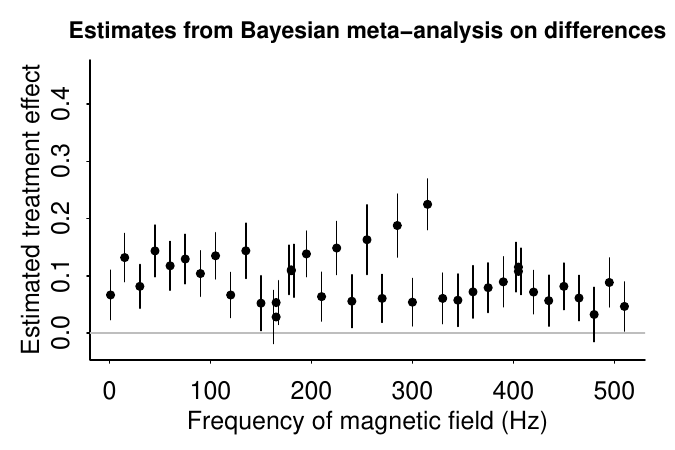}}
  \vspace{-.1in}
  \caption{\em (a) Posterior mean $\pm$ standard deviation of each treatment
  effect $\theta_j$ from our proposed hierarchical model that separately models 
  treatment and sham effects. (b)
  For comparison, the estimates given by a Bayesian meta-analysis in the sense of a 
  hierarchical model for the difference estimates.}
  \label{bayesian-metaanalysis}
\end{figure}

\section{Alternative simulation study based on raw estimates}
\label{alt-sim}

\begin{figure}
  \centerline{\includegraphics[width=.9\textwidth]{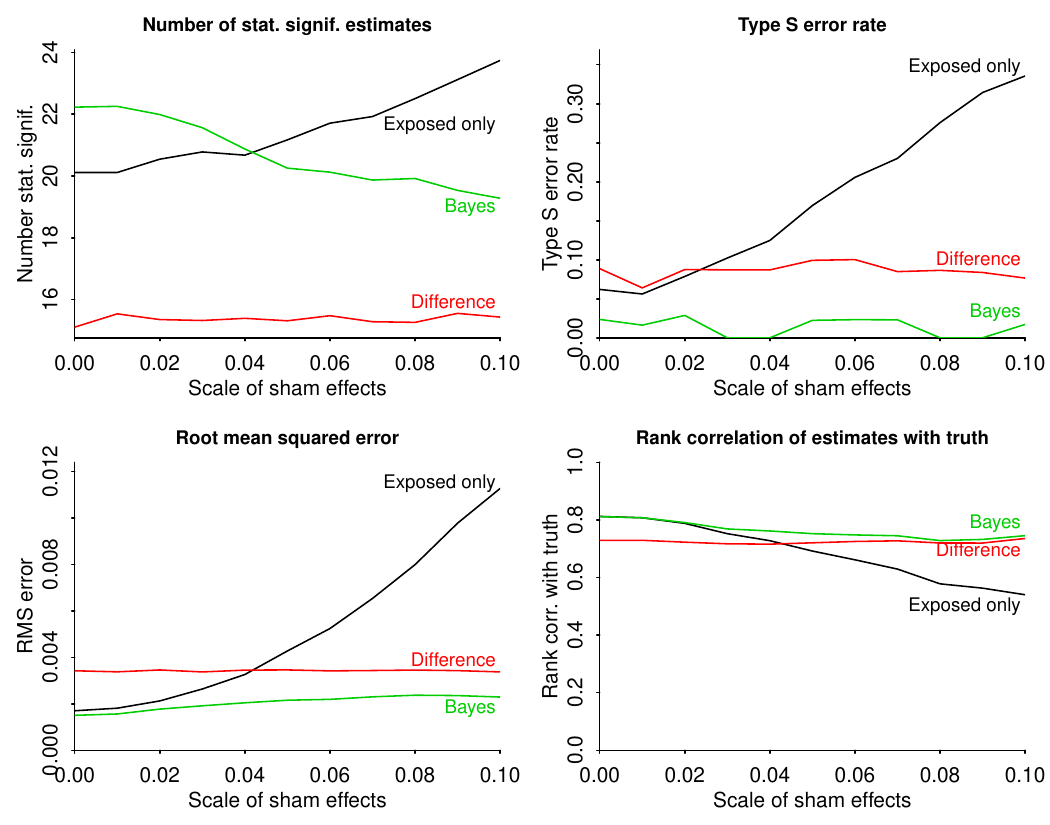}}
  \vspace{-.1in}
  \caption{\em Results of the alternative simulation study from Appendix
  \ref{alt-sim} comparing three estimates---the exposed data estimate,
  $y_{j1}$, the difference between exposed and sham, $y_{j1}-y_{j0}$, and the
  Bayesian hierarchical model estimate $\mbox{E}(\theta_j|y)$---to simulated
  data.  The four graphs show the results for four different frequency
  evaluations, and on each graph the horizontal axis represents $\sigma^b$, the
  standard deviation of the sham effects in the simulation.}
  \label{simulation-alt}
\end{figure}

One objection the reader might have to our simulation study of Section
\ref{subsec:simulation} is that we were simulating the $\theta_j$ from the
posterior as obtained from our own Bayesian model, which might give the Bayesian estimates
an unfair advantage. In this section, to address this concern, we show that we
observe the same phenomena as discussed in Section \ref{subsec:simulation} even
if we use the raw estimates for $\theta_j$ instead.

Specifically, we perform this alternative simulation exactly as before described except that we perform the following two steps
instead of steps 1--3 in
Section \ref{subsec:simulation}: (1) Simulate one draw of the vector of 38 values $b_j,
j=1,\dots,J$, drawing them independently from the $\mbox{normal}(0,\sigma^b)$
distribution; (2) Simulate one dataset, that is a vector of 38 values
$y_{j0}\sim {t}_{n_{j0}-1}(b_j, \sigma^y)$ and a vector of 38 values $y_{j1}\sim
{t}_{n_{j1}-1}(y_{1j}^{obs}+b_j, \sigma^y)$, where we write $y_{j1}^{obs}$ for
the raw exposed-only estimates of the treatment effects from the actual chick
data as observed by Blackman et al.\cite{BlackmanBenaneElliottHousePollock1988}.  We are thus centering our estimated
treatment effects at the observed data rather than, as before, at the hierarchical Bayes estimates.

The results are summarized in Figure \ref{simulation-alt}. They tell mostly the
same story as we saw in our original simulation study. One difference is that
the exposed-only estimate now consistently results in more statistically significant
estimates compared to the Bayesian estimate. However, inspection of the
type S error rates reveals that these extra significant estimates are not to be
trusted. The results of this simulation show that the hierarchical Bayesian estimate is still
superior to its two alternatives even in cases where its partial partial pooling
behavior is not an advantage.

\section{Simulation study for differently sized datasets}\label{simulation-sizes}
The reader might wonder how many repeated experiments are needed for our method
to be superior, in a practical sense, compared to existing default analyses like 
difference or plain exposed estimates.
To give a partial answer to this question, we repeat the simulation study of Figure \ref{simulation}
here, for datasets of varying sizes $M=5,10,20,38$.

To be precise, we set $\mu^b$ to 0 and consider a range of values for $\sigma^b$, for each
performing the following steps 200 times:  (1) Simulate one draw of the vector
of $M$ values $b_j, j=1,\dots,M$, drawing them independently from the
$\mbox{normal}(0,\sigma^b)$ distribution; (2) Draw the vector of the $M$ values
$\theta_j,j=1,\dots,M$, from their (joint) posterior distribution from Section
\ref{posterior.fit}; (3) Simulate one dataset, that is a vector of $M$ values
$y_{j0}\sim \mbox{normal}(b_j, \sigma^y)$ and a vector of $M$ values $y_{j1}\sim
\mbox{normal} (\theta_j+b_j, \sigma^y)$.

For the purposes of reducing computational instability that might arise for small datasets, 
we add weakly informative priors to our model of Section \ref{bayes} in the above simulation.
To be precise, we add a $\mbox{normal}(0,1)$ prior to both $\mu^\theta$ and $\mu^b$ and a $\mbox{half-normal}(0,1)$
prior to both $\sigma^\theta$ and $\sigma^b$.
These priors are strong enough to rule out unreasonable parameter values and improve the computational behavior 
of the model for small datasets, but they are weak enough not to materially affect the resulting posterior estimates.

\begin{figure}
  \begin{tabular}{lll}
  {\includegraphics[width=.46\textwidth]{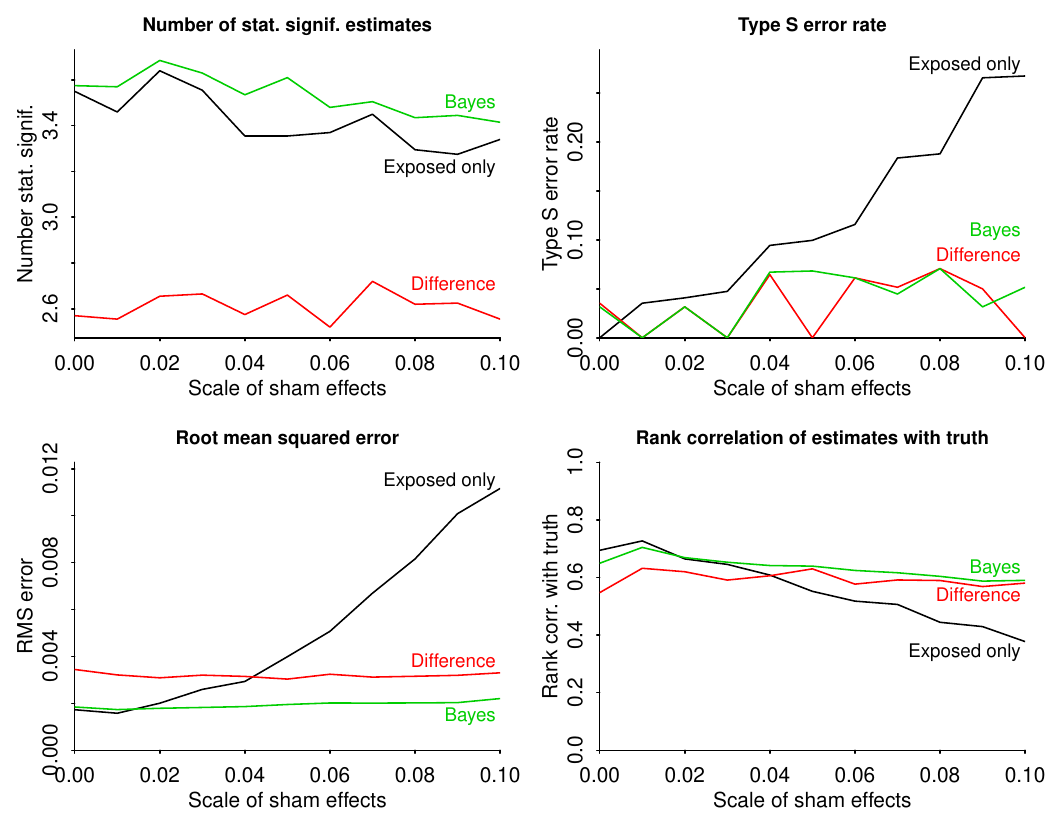}}&&
   {\includegraphics[width=.46\textwidth]{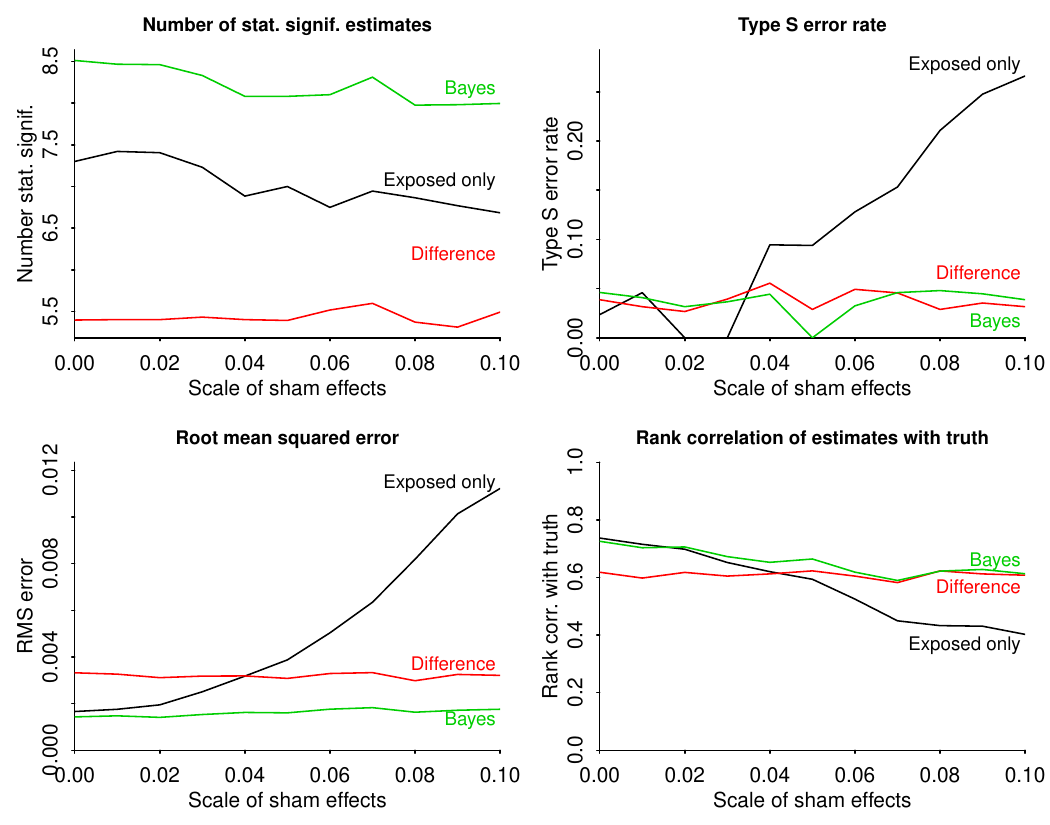}}\\
 & &\\
  {\includegraphics[width=.46\textwidth]{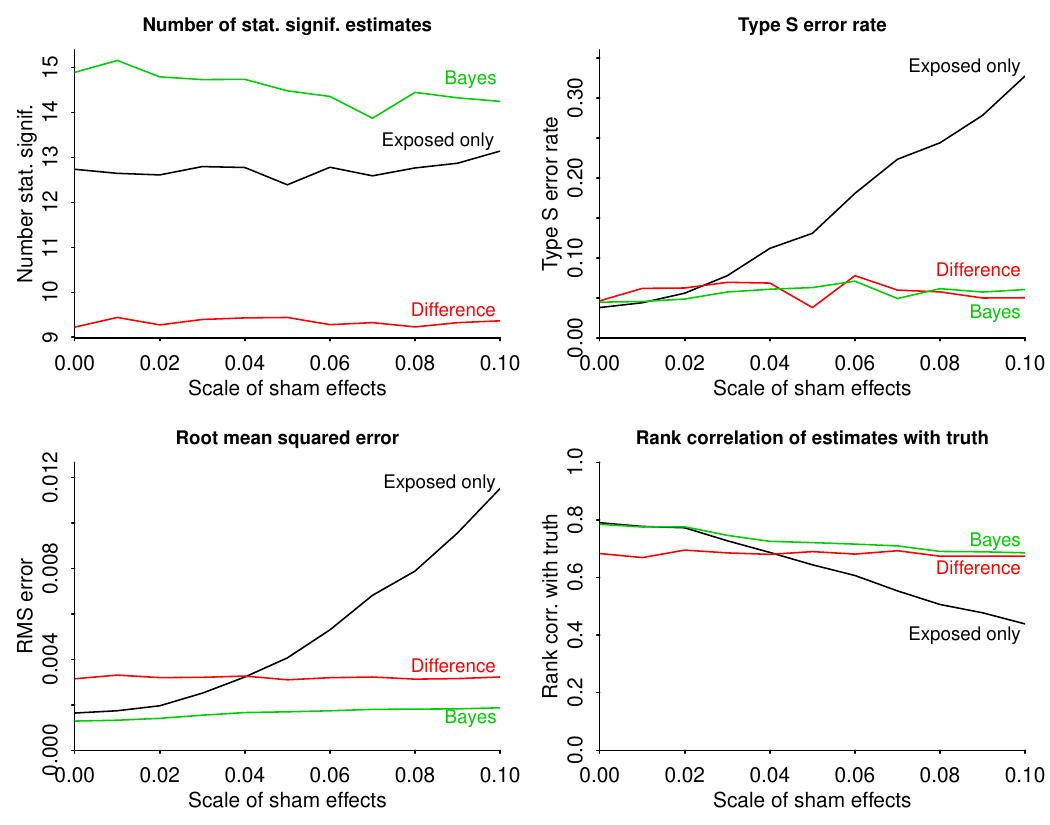}}&&
  {\includegraphics[width=.46\textwidth]{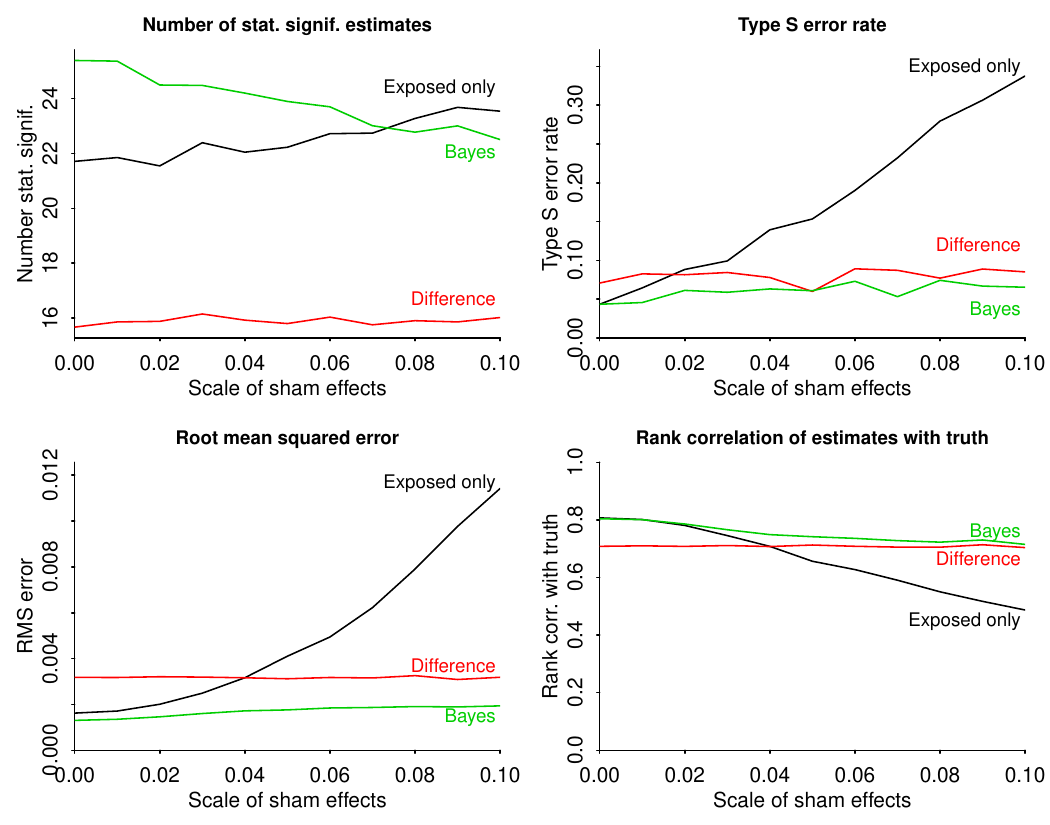}}
  \end{tabular}
  \vspace{-.15in}
  \caption{\em Results of simulation studies for datasets of 
   sizes 5 (top left), 10 (top right), 20 (bottom left), and 38 (bottom right), comparing three estimates---the
  exposed data estimate, $y_{j1}$, the difference between exposed and sham,
  $y_{j1}-y_{j0}$, and the Bayesian hierarchical model estimate
  $\mbox{E}(\theta_j|y)$---to simulated data.  The four graphs show the results
  for four different frequency evaluations, and on each graph the horizontal
  axis represents $\sigma^b$, the standard deviation of the sham effects in the
  simulation.}
  \label{simulation2}
\end{figure}

We present the results of the simulation studies in Figure \ref{simulation2}.
We see that our method always outperforms the existing default analyses according to our four metrics,
on our particular example simulated datasets.
Further, we see that the gained performance does not seem to decrease notably as the number of repeated experiments get decreased down to 5.  The noisiness of some of the graphs in Figure \ref{simulation2} could be reduced by increasing the number of simulations, but this is not necessary here because the general pattern is clear.

What happens for even smaller numbers of repeated experiments, we leave to future research.
In those scenarios, we expect that users would do well to augment their models with stronger priors on the hyperparameters. 

\section{Alternative model for example 2}\label{alt-mod2}\label{alt-mod-binomial}
In this appendix, we discuss an alternative model we could have used for
analyzing the data of example 2 in Section \ref{berlim}.  In Section \ref{berlim} we used the hierarchical normal model using a standard correction for zero counts.  But, given that we fit our models in Stan, we could just have easily have modeled the discrete data more directly, using a binomial likelihood rather than a normal approximation.
This leads to the following model, directly modeling $n_{ji}$, rather than $y_{ji}$:

\begin{figure}
  \centerline{
    \includegraphics[width=.5\textwidth]{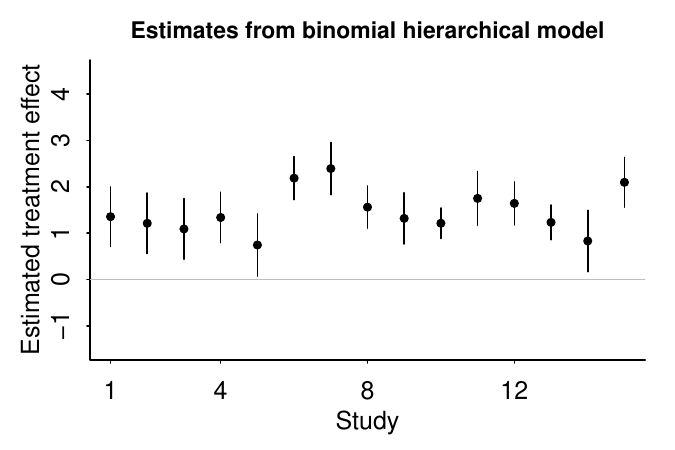}
    \includegraphics[width=.5\textwidth]{berlim4a.pdf}}
  \vspace{-.15in}
  \caption{\em (a) Posterior mean $\pm$ standard deviation of each treatment effect $\theta_j$ from the hierarchical model (\ref{binomial-eq}) with binomial likelihood. (b)
  For comparison, the same estimates from the hierarchical model (\ref{meas}) with normal likelihood.}
  \label{binomial}
\end{figure}

\begin{eqnarray}
  \nonumber  b_j &\sim&\mbox{normal} (\mu^{b}, \sigma^{b} )\\
  \nonumber      \theta_j &\sim&\mbox{normal} (\mu^{\theta}, \sigma^{\theta} )\\
  \nonumber n_{j0} &\sim & \mbox{binomial}(N_{j0}, \mbox{logit}^{-1}(b_j))\\
  \label{binomial-eq} n_{j1} & \sim & \mbox{binomial}(N_{j1}, \mbox{logit}^{-1}(\theta_j+b_j)).
\end{eqnarray}
Figure \ref{binomial} shows how this leads to largely the same conclusions, but with slightly less pooling and higher uncertainty than the hierarchical model with the normal likelihood function.

\begin{table}
  \begin{small}
    \centerline{
      \begin{tabular}{crr}
        Parameter          & Estimate (s.e.) & 95\% interval        \\\hline
        $\mu^{\theta}$     & 1.5 (0.3)   & $\ \ [0.8, 2.2]$ \\
        $\sigma^{\theta} $ & 0.7 (0.3)   & $\ \ [0.2, 1.4]$ \\
        $\mu^{b}$          & $-$3.0 (0.2)   & $[-3.5, -2.5]$    \\
        $\sigma^b$         & 0.3 (0.2)   & $\ \ [0.0, 0.7]$ \\
      \end{tabular}
    }
  \end{small}
  \caption{\em Posterior means, standard deviations, and 95\% intervals  for the
    hyperparameters in the binomial hierarchical model fit to the rTMS
    data.}\label{berlim-inferences-binomial}
\end{table}

We summarize the estimated hyperparameters in Table \ref{berlim-inferences-binomial}, which shows
slight differences from the estimates of Table \ref{berlim-inferences}.
The estimates of $\mu^{\theta}$ and $\sigma^{\theta}$ have increased while those for $\mu^{b}$ and $\sigma^{b}$ have decreased in the binomial model.
This results in lower odds of remission for patients receiving the sham treatment as well as a larger relative effect of the real treatment compared to the sham treatment.
Moreover, the binomial model reports larger standard errors for all hyperparameters.

\end{document}